\documentclass[prb,twocolumn,superscriptaddress,longbibliography,aps]{revtex4-1}

\usepackage[sumlimits]{amsmath} 
\usepackage{amssymb}
\usepackage{bm}
\usepackage{graphicx}
\usepackage{epstopdf}
\usepackage{latexsym}
\usepackage{subfigure}
\usepackage{color}
\usepackage{nicefrac}
\usepackage{dsfont} 
\usepackage{wasysym} 
\usepackage{stmaryrd} 
\usepackage{hyperref} 
\usepackage{xcolor}
\usepackage{empheq} 
\usepackage{relsize}

\usepackage{sansmath}
\DeclareFontEncoding{LGR}{}{}
\DeclareSymbolFont{sfgreek}{LGR}{cmss}{m}{n}
\SetSymbolFont{sfgreek}{bold}{LGR}{cmss}{bx}{n}
\DeclareMathSymbol{\sxi}{\mathord}{sfgreek}{`x}
\DeclareMathSymbol{\stheta}{\mathord}{sfgreek}{`j}
\DeclareMathSymbol{\salpha}{\mathord}{sfgreek}{`a}
\DeclareMathSymbol{\ssigma}{\mathord}{sfgreek}{`s}
\DeclareMathSymbol{\sepsilon}{\mathord}{sfgreek}{`e}
\DeclareMathSymbol{\sOmega}{\mathalpha}{sfgreek}{`W}
\DeclareMathSymbol{\sPhi}{\mathord}{sfgreek}{`F}
\DeclareMathSymbol{\sPsi}{\mathord}{sfgreek}{`Y}

\usepackage{verbatim}
\usepackage{caption}
\usepackage{cancel}
\usepackage{url}
\usepackage{soul} 

\newcommand{\mb}{\mathbf}
\newcommand{\bs}{\boldsymbol}

\newcommand{\mc}{\mathcal}
\newcommand{\ms}{\mathsf}
\newcommand{\vd}{{\vphantom \dagger}}

\renewcommand{\d}{\text d}

\usepackage{subcaption}
\usepackage{ragged2e}
\DeclareCaptionJustification{justified}{\justifying}
\begin{document}

\title{Anomalous quantum oscillations from boson-mediated interband scattering}

\author{L\'eo Mangeolle}
\affiliation{Technical University of Munich, TUM School of Natural Sciences, Physics Department, 85748 Garching, Germany}
\affiliation{Munich Center for Quantum Science and Technology (MCQST), Schellingstr. 4, 80799 M{\"u}nchen, Germany}
\author{Johannes Knolle}
\affiliation{Technical University of Munich, TUM School of Natural Sciences, Physics Department, 85748 Garching, Germany}
\affiliation{Munich Center for Quantum Science and Technology (MCQST), Schellingstr. 4, 80799 M{\"u}nchen, Germany}
\affiliation{Blackett Laboratory, Imperial College London, London SW7 2AZ, United Kingdom}

\date{\today}
\begin{abstract}
Quantum oscillations (QO) in metals refer to the periodic variation of thermodynamic and transport properties as a function of inverse applied magnetic field. QO frequencies are normally associated with semi-classical trajectories of Fermi surface orbits but recent experiments challenge the canonical description. We develop a theory of composite frequency quantum oscillations (CFQO) in two-dimensional Fermi liquids with several Fermi surfaces and interband scattering mediated by a dynamical boson, e.g. phonons or spin fluctuations. Specifically, we show that CFQO arise from oscillations in the fermionic self-energy with anomalous frequency splitting and distinct strongly non-Lifshitz-Kosevich temperature dependencies. Our theory goes beyond the framework of semi-classical Fermi surface trajectories highlighting the role of  many-body effects. We provide experimental predictions and discuss the effect of non-equilibrium boson occupation in driven systems. 
\end{abstract}

\maketitle

\textit{Introduction:}
Magneto-oscillations -- or ``quantum'' oscillations (QO) -- have been instrumental in probing Fermi surfaces since their discovery in 1930. 
  They appear in both thermodynamic quantities such as the magnetic susceptibility -- the de Haas - van Alphen (dHvA) oscillations \cite{de1930dependence} --
  and transport properties such as the electric conductivity -- the Shubnikov - de Haas (SdH) oscillations \cite{schubnikow1930magnetic}.
  The semiclassical theory developed by Onsager \cite{onsager1952interpretation} and the subsequent calculations by Lifshitz and Kosevich \cite{lifshitz1956theory}
  relate the fundamental frequencies appearing in QO experiments, as well as the temperature dependence of their amplitude -- the Lifshitz-Kosevich (LK) behavior --,
  to geometric properties of the Fermi surface (FS). Thus, QO allow for the precise determination of FSs~\cite{pippard1960experimental,shoenberg1984magnetic} and the inclusion of Berry phase effects in its semi-classical description  established the field of ``fermiology''~\cite{alexandradinata2018revealing,alexandradinata2023fermiology}.

Other QO frequencies in addition to the fundamental ones often appear in magnetic breakdown situations \cite{fischbeck1970theory,alexandradinata2018semiclassical},
  and are proportional to the extremal enclosed area of tunneling-allowed semi-classical orbits. 
  Their temperature dependence still follows the usual LK dependence, $R_{\rm LK}^{[m]} (T) \equiv \frac{2\pi^2 T m}{\sinh(2\pi^2Tm)}$,  
  with an effective mass $m$ consistent with the semi-classical picture. 
  However, recent experiments~\cite{huber2023quantum,phinney2021strong} have observed unusual CFQO frequencies set by the difference of two fundamental frequencies not connected via magnetic breakdown. These \emph{difference frequency} quantum oscillations (DFQO)~\cite{leeb2023theory} do not have an intuitive semi-classical picture \`a la Onsager and are, in that sense,
a purely quantum effect induced by scattering between distinct FSs. 
Their temperature dependence follows the LK law with a typical temperature set by the difference between the masses of the two FSs involved in the scattering, allowing them to survive thermal smearing to much higher temperatures than the fundamental frequency QOs. 
More broadly, CFQO can be thought of as generalizations of magneto-intersubband QO~\cite{raikh1994magnetointersubband,grigoriev2003theory} known in coupled 2DEGs and quasi-two-dimensional materials, which may also arise from inter-band interactions~\cite{allocca2021low}. 
Remarkably, a whole range of known (bulk) metals possibly display CFQO, see Ref.~\cite{leeb2024field} for a recent review, and other correlated quantum materials with unusual QO temperature dependence~\cite{mccollam2005anomalous,tan2015unconventional,galeski2024quantum} have been reported. The complex QO spectrum of FeSe displays possible DFQO~\cite{leeb2024interband} which could originate from two FSs connected by strong spin fluctuation scattering observed experimentally~\cite{wang2016magnetic} calling for a better understanding of many-body effects.

\begin{figure*}[!t]
\centering
\includegraphics[width=\textwidth]{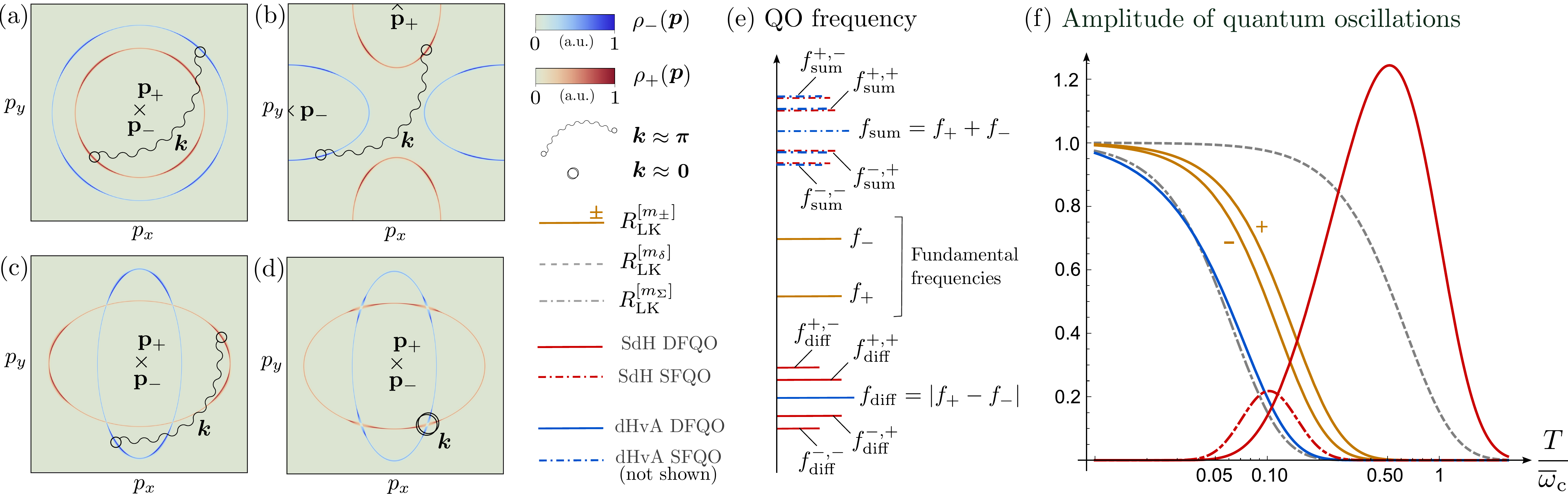}
    \caption{(a-d) The different Fermi surface types described by our effective model including schematic scattering with boson wavevectors $\bs k$. Color scale: the dressed fermionic spectral function (including fermion-boson self-energy) at the Fermi level $\rho_s(\bs p)= - \frac 1 \pi {\rm Im} G_{s,s}(\bs p,0)$. 
      (e) Summary of the fundamental frequencies $f_\pm=\mu_\pm m_\pm/\sf e$ and main composite frequencies appearing in the Fourier transform of an experimental quantity as a function of $1/B$ (not to scale), for $\mu_+=\mu_-$. We show the splittings with $\delta\!f_s \equiv  \bar \Omega_\lambda m_s/\sf e$, resulting in several peaks at $f_{\rm sum}^{\pm,s} \equiv f_{\rm sum}\pm \delta\!f_s$ and $f_{\rm diff}^{\pm,s} \equiv f_{\rm diff}\pm \delta\!f_s$. The height of the peaks depicted is arbitrary.
      (f) Temperature dependence of various QO amplitudes, including SdH DFQO ${\sf A}_{\sigma_{\rm L}}^{s,\lambda}$ and SdH sum frequency quantum oscillations (SFQO) ${\sf A}_{\sigma_{\rm L},{\rm sf}}^{s,\lambda}$ in the same units and dHvA DFQO $\hat {\sf A}_{n_{\rm p}}^{s,\lambda}$. Units, notations and parameter values are those of Fig.~\ref{fig:plots} with $\bar \Omega_{\lambda}=0.5$.
      The fundamental LK functions $R_{\rm LK}^{[m_\pm]}(T)$ and the CFQO LK functions $R_{\rm LK}^{[m_\delta]}(T),R_{\rm LK}^{[m_\Sigma]}(T)$,
      where $m_\Sigma = m_++m_-$ and $m_\delta = |m_+-m_-|$, are plotted for comparison.}
\label{fig:advertisement}
\end{figure*}

In this paper, we investigate QO in two-dimensional (2D) systems with two ellipse-shaped Fermi surfaces and interband scattering mediated by a dynamical boson.
Our generic model applies to a whole range of different systems, see Fig.\ref{fig:advertisement} panels (a--d), for example to parent compounds of bilayer high-T$_c$ superconductors~\cite{garcia2010multiple}, metals with split bands from spin orbit coupling~\cite{manchon2015new} like CoSi~\cite{huber2023quantum} or MoSi$_2$~\cite{pavlosiuk2022giant}, $d$-wave altermagnets~\cite{vsmejkal2022emerging}, or parent compounds of iron based superconductors like FeSe~\cite{wang2016magnetic,leeb2024interband}. The boson could originate from spin fluctuations connecting hot spots on the Fermi surface~\cite{abanov2003quantum}, simply from electron-phonon interactions or other collective modes. 
We study the coupled electron-boson problem by means of diagrammatic many-body theory in a magnetic field \cite{fowler1965electron, engelsberg1970influence} assuming the validity of Fermi liquid theory~\cite{luttinger1961theory, gor1962quantum, wasserman1996influence}.

 We show that CFQO without a semi-classical origin are generated from dynamical boson-fermion scattering, and that the boson dynamics induces a frequency splitting, which is notably absent in the fundamental frequency case~\cite{luttinger1961theory, wasserman1996influence}, see  Fig.\ref{fig:advertisement}(e) for a summary of the frequencies. 
 Due to the thermal activation of bosons, the CFQO temperature dependences display strongly non-LK behavior, see Fig.\ref{fig:advertisement}(f), which evades the Fowler-Prange theorem constraining the temperature dependence of fundamental frequency QOs in the presence of interactions~\cite{fowler1965electron, engelsberg1970influence, wasserman1996influence, martin2003quantum}. These features, both anomalous, are allowed for composite frequencies while not for fundamental frequencies, and are consequences of interband scattering mediated by a \emph{dynamical} excitation. Indeed, they are absent in the recently discussed disorder-induced CFQO~\cite{leeb2023theory,leeb2024interband}, and in the static case where the boson condenses. The latter corresponds to a density wave phase in which the fermion-boson interaction at the saddle-point reduces to a band hybridization, which leads to a magnetic breakdown situation~\cite{fischbeck1970theory} obeying the LK paradigm and without difference frequency oscillations.

\textit{Model and Setup:}
We consider a two-dimensional system with two bands of electrons or holes, indexed by a quantum number $s=\pm$, with anisotropic quadratic dispersion relations, in a magnetic field. 
The index $s$ can refer to any quantum number like spin, orbital, valley, etc. The minimal coupling Hamiltonian in the Landau gauge $A_x=-yB,A_y=0$, where $\bs A$ is the vector potential and $B$ is the magnetic field, reads
\begin{align}
  \label{eq:main1}
  H_0 &= \textstyle{\sum_{\bs p}\sum_{s = \pm}}  \; | \bs p,s \rangle \Big [ \frac 1 {2m_s} \left ( \alpha_s^2 (p_x-\text p_s^x-y)^2 \right . \nonumber \\
  & \left . \qquad + \alpha_s^{-2} (p_y-\text p_s^y)^2 \right ) \vphantom{\frac 1 {2m_s}} - \mu_s \Big ]  \langle \bs p,s |,
\end{align}
where $\bs p$ stands for momentum, and $m_s$, $\mu_s$, $\alpha_s$ and $\mb p_s$ are the band-dependent mass, chemical potential,
anisotropy factor and position in $\bs p$-space of the center of the dispersion of band $s$, respectively. In Fig.\ref{fig:advertisement} panel (a-d) we plot example Fermi surfaces covering a whole range of materials discussed above.
Throughout the paper, to focus on the salient harmonic content of QOs we fix the magnetic length $\ell_{\rm c} = ({\sf e}B)^{-1/2}$, where $-\sf e$ is the electron charge, to unity, $\ell_{\rm c}=1$\footnote{Note, then, that $\omega_{{\rm c},s}=1/m_s$.}; besides we use units where $\hbar = 1 = k_{\rm B}$.
  
The free hamiltonian can be easily diagonalized into $H_0 = \sum_{s,n,p_x} \varepsilon_{s,n}  |s,n, p_x \rangle \langle s,n, p_x |$,
  where the energies are $\varepsilon_{s,n} = \omega_{{\rm c}, s} \big ( n + \tfrac 1 2 \big ) - \mu_s $, with $\omega_{{\rm c}, s} := \frac{{\sf e} B}{m_s}$ the cyclotron frequency,
  and the normalized eigenstates $ |s,n, p_x \rangle  $ are indexed by quantum number $s$, level index $n$, and degenerate momentum $p_x \in \llbracket 0 ; \frac{2\pi}{L_x} ; \cdots ; L_y \rrbracket $
  yielding the usual Landau level degeneracy $N_\phi = \frac{ L_x L_y}{2\pi}$ ~\cite{landau1930diamagnetismus}.
  The free fermion propagator $G_0(z)= \frac{1}{z - H_0}$, a function of $z$ the complex frequency, is diagonal in both the $s$ and $n$ indices~\cite{mahan1984electron}.
  We provide its explicit real-space expression in the SM~\cite{supp}.
  
\begin{figure}[htbp]
  \begin{center}
    \includegraphics[width=.9\columnwidth]{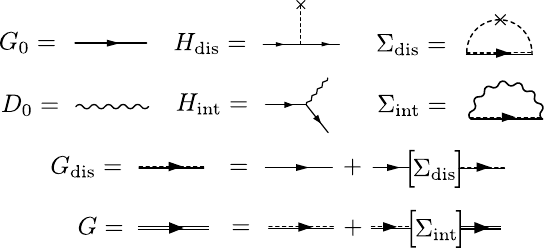}
\caption{Diagrammatic summary of the setup and notations for interactions, propagators and self-energies.}
\label{fig:diagrams}
\end{center}
\end{figure}

  We include the effect of a local scalar disorder potential $H_{\rm dis} = \sum_{s,\mb r} u(\mb r)  c^\dagger_{s,\mb r}  c^{\vphantom \dagger}_{s,\mb r} $,
  where $c_{s,\mb r}$ is the local fermion destruction operator at position $\mb r$, and we assume that the disorder potential
  is uncorrelated: $\langle u(\mb r) u(\mb r') \rangle = \bar u_0^2 \delta(\mb r - \mb r')$ with $\bar u_0$ a constant.
  Such disorder does not carry any quantum number $s$, therefore the broadened fermionic propagator $G_{\rm dis}(z)=\frac{1}{G_0(z)^{-1}-\Sigma_{\rm dis}(z)}$  remains diagonal in $s$.
  The disorder self-energy, which we compute in the non-crossing approximation (see Fig.\ref{fig:diagrams})
  as $\Sigma_{\rm dis}(\bs r,\bs r',z) = \bar u_0^2 \delta(\bs r - \bs r') G_{\rm dis}(\bs r,\bs r',z)$,
  can be determined self-consistently and reads, in the basis of Landau eigenstates,
 $(\Sigma_{\rm dis})_{s,s'}^{n,n'}(p_x,p_x';z)   = - i \Gamma_{s}^{\rm dis}\, {\rm sign}({\rm Im}(z)) \, \delta_{s,s'} \delta_{n,n'} \tfrac{2\pi}{L_x}\delta(p_x-p_x') $,
  with scattering rate $\Gamma_{s}^{\rm dis}\equiv  \frac{\overline u_0^2}{ 2\omega_{{\rm c},s}}$~\cite{abrikosov1975methods}.
  This leads to a Dingle factor  $R_{\rm D}^s \equiv \exp ( -2\pi  \Gamma_{s}^{\rm dis}/ \omega_{s,\rm c} )$ which damps out any magneto-oscillating terms,
  and thus serves as a small parameter controlling expansions in higher harmonics.
  
Next we consider an interaction between the fermions $c^\dagger_{s,\bs p}  $ and a neutral boson $a^\dagger_{\lambda,\bs k} $ carrying one unit of the quantum number $s$
and possibly a flavor index $\lambda$, with dispersion $\Omega_{\lambda,\bs k}$ specified below.
The boson mediates an interband scattering between the two Fermi surfaces but does not allow for intraband scattering.
The interaction vertex in zero field reads
\begin{align}
  \label{eq:main47}
   H_{\rm int} &= \frac 1 {\sqrt N} \textstyle{\sum_{\lambda,\bs k}\sum_{s,\bs p}} (2\Omega_{\lambda,\bs k})^{-1/2} g_{s,\overline s}^{\lambda,\bs k} \nonumber \\
     &\qquad \times \, c^\dagger_{\overline s,\bs p+\bs k}  \, \big ( a^\vd_{\lambda,\bs k} + a^\dagger_{\lambda,-\bs k} \big )  \, c^{\vphantom \dagger}_{s,\bs p} ,
\end{align}
where $\overline s \equiv -s$, $N$ is the total number of unit cells, 
hermiticity imposes the relation $g_{s,\overline s}^{\lambda,\bs k} = ( g_{\overline s,s}^{\lambda,-\bs k} )^*$,
and the absence of any $\bs p$ dependence in the coupling constant $g_{s,\overline s}^{\lambda,\bs k} $ is due to gauge invariance
and the fact that the boson is neutral.

\textit{Boson-mediated self-energy oscillations:}
Physical quantities are functions of the full interacting fermionic propagator,
$G(z) = \frac 1 {G_{\rm dis}^{-1} - \Sigma_{\rm int}(z)}$. We compute the fermion-boson self-energy $\Sigma_{\rm int}(z)$ 
assuming the interaction strength $g$ to be small to the order of one loop, i.e. $O(g^2)$. We obtain, in the Landau level basis, 
\begin{align}
  \label{eq:main50}
   &(\Sigma_{\rm int})_{s,s'}^{n,n'}(z) = \delta_{s,s'}    \sum_{l=0}^\infty   \int_{\bs k}   \sum_{\lambda} \frac{2\pi}{\Omega_{\lambda,\bs k}}\,\left | g_{s,\overline s}^{\lambda,\bs k} \right |^2 \, I_{s,l}^{n,n'}(\bs k) \nonumber \\
  &\quad \times  \left [  \frac{n_{\rm B}(\Omega_{\lambda,\bs k}) + n_{\rm  F}( \tilde \varepsilon_{\overline s,l}   )  }{z +\Omega_{\lambda,\bs k} - \tilde \varepsilon_{\overline s,l} }
                                     - ( \Omega_{\lambda,\bs k} \leftrightarrow - \Omega_{\lambda,\bs k} ) \right ] ,
\end{align}
with $l$ an internal Landau level index and the degeneracy $\frac{2\pi}{L_x}\delta(p_x-p_x')$ is implicit.
Here $ \int_{\bs k} \equiv \int \tfrac{\text d^2 \bs k}{(2\pi)^2}$,
$n_{\rm B}(\omega)=\frac 1 {e^{\omega/ T}-1}$ is the Bose function,
$n_{\rm F}(\omega)=\frac 1 {e^{\omega/ T}+1}$ is the Fermi function,
$\tilde \varepsilon_{s,l} \equiv \varepsilon_{ s,l} - i \Gamma_{ s}^{\rm dis}\, {\rm sign}({\rm Im}(z)) $, 
and $I_{s,l}^{n,n'} (\bs k)$ is some static matrix element in the Landau eigenbasis~\cite{supp, bateman1953higher}.
Because the self-energy $(\Sigma_{\rm int})_{s,s}^{n,n'}(z)$ for band $s$ contains an internal summation over Landau levels $l$ of the \emph{other} band $\overline s$, with energies $\varepsilon_{\overline s,l}$,
it displays magneto-oscillations at frequencies determined by the $\overline s$ Fermi surface and by the boson dispersion. We stress that this is a dynamical effect of electron-boson scattering, arising from the energy dependence in the denominator of Eq.\eqref{eq:main50}. Also note that magneto-oscillations of the self-energy can be especially large in 2D~\cite{martin2003quantum, bychkov1983two}.

For these to exist, the non-conservation of the Landau level index upon scattering is crucial. 
Indeed, there are as many such oscillations as Landau levels $l$ whose contribution to the sum in Eq.\eqref{eq:main50} is sizeable. 
This, in turn, depends strongly on $I_{s,l}^{n,n'}(\bs k)$ which puts constraints on the models allowing for these oscillations.
With our choice Eq.\eqref{eq:main1}, we checked that sufficiently different anisotropic fermion dispersions ($\alpha_+ \neq \alpha_-$),
or bosons dispersing around a vector $\bs Q\neq \bs 0$ (for instance $\bs Q = \bs \pi$ the N\'eel vector), allow for a large number of oscillations in the self-energy.
Besides, the absolute amplitude of the latter is larger when the self-energy itself is larger,
which is favored by ``crossing'' or ``nesting'' configurations \cite{chan1973spin}, depending on the boson's dispersion as illustrated with the color scale in Fig.\ref{fig:advertisement}(a-d).

As magneto-oscillations are mainly determined by electronic properties close to the quasiparticle peak~\cite{wasserman1996influence},
we study oscillations of $\Sigma_n^s(\epsilon)\equiv (\Sigma_{\rm int})_{s,s}^{n,n}(\epsilon)$ by focusing on $\epsilon \approx \varepsilon_{s,n}$, i.e.\  when $n$ takes the value $\xi_s(\epsilon) = (\epsilon+\mu_{s})/\omega_{s,\rm c}$.
These oscillations can be made explicit by Poisson-resumming Eq.\eqref{eq:main50}, and the resulting expression has the form~\cite{supp}
\begin{align}
  \label{eq:main61}
  & \Sigma_{\xi_s(\epsilon)}^s(\epsilon)\approx {\sf \Sigma}_s(\epsilon) \\
  &\quad + \textstyle{\sum_\lambda\sum_{\eta=\pm}}i ({\sf \Sigma}')_{s,\overline s}^{\lambda,\eta}(\epsilon)
    \,    e^{i 2\pi \left [ \xi_{\overline s}(\epsilon) + \eta \frac {\bar{\Omega}_\lambda}{\omega_{{\rm c},\overline s}} \right ] }\nonumber \\
  &\quad   + \textstyle{\sum_\lambda\sum_{\eta=\pm}} ({\sf \Sigma}'')_{s,\overline s}^{\lambda,\eta}(\epsilon)
                                    \,   \sin \left ( 2\pi \xi_{\overline s}(0) \right ) \;+  \dots, \nonumber
\end{align}
where $ {\sf \Sigma}_s$, $({\sf \Sigma}')_{s,\overline s}^{\lambda,\eta}$  and $({\sf \Sigma}'')_{s,\overline s}^{\lambda,\eta}$ are slowly varying complex functions of $\epsilon$,
oscillating terms come with Dingle factors
(in particular $({\sf \Sigma}')_{s,\overline s}^{\lambda,\eta}\propto R_{\rm D}^{\overline s} $, $({\sf \Sigma}'')_{s,\overline s}^{\lambda,\eta}\propto R_{\rm D}^{\overline s} $),
and ``$+\dots$'' contains terms which are higher order in Dingle factors (i.e.\ higher harmonics).
In ``$+\dots$'' we also included the first harmonic term, arising from impurity scattering, which oscillates at the frequency of the $s$ band:
this term does not contribute to CFQOs to the first order in Dingle factors, and is thus irrelevant here.

In principle, any boson energy $\Omega_{\lambda,\bs k}$ for any $\bs k$ yields a different frequency at which $ \Sigma_{\xi_s(\epsilon)}^s(\epsilon) $ oscillates.
The spectrum of $\Sigma_{\xi_s(\epsilon)}^s(\epsilon)$ as a function of $1/B$ thus displays a continuous distribution of frequencies
$(m_{\overline s}/{\sf e}) \left [ \epsilon + \mu_{\overline s} \pm \Omega_{\lambda,\bs k} \right ]$ centered around its value at $\Omega_{\lambda,\bs k} \rightarrow \bar{\Omega}_\lambda$,
where $\bar{\Omega}_\lambda(T)$ denotes an averaged bosonic frequency.
In Eq.\eqref{eq:main61}, for clarity we gathered all contributions into a single oscillating term at the average frequency
$(m_{\overline s}/{\sf e})\left [ \epsilon + \mu_{\overline s} \pm \bar{\Omega}_\lambda \right ]$. We note that the assumption, which we make also in the following, is exact for a flat bosonic dispersion.

\textit{Difference frequency SdH and dHvA effects:}
We will now focus on those contributions to the electrical conductivity $\sigma_{xx}$ and the particle density $n_{\rm p}$ which oscillate with $1/B$
at the difference frequency $f_{s,\overline s} \equiv  \left ( m_s \mu_{s} - m_{\overline s}\mu_{\overline s} \right )/{\sf e} $, or closeby. 
\footnote{While CFQO can be frequency-split, the fundamental frequencies and their higher harmonics strictly retain their semiclassical values -- for \emph{purely interband} boson-mediated scattering and to the order we consider.}

We compute DFQO in the longitudinal conductivity \cite{bastin1971quantum}
\begin{align}
  \label{eq:main62}
  \sigma_{\rm L}&= - \frac{{\sf e}^2/\pi}{L_x L_y}\int \text d \epsilon
               \frac{\mb {Tr}\left [ \hat v_x \,{\rm Im} G(\epsilon) \,\hat v_x \,{\rm Im} G(\epsilon) \right ]}{4T \cosh^2 \left ( \frac{\epsilon }{2T} \right )},
\end{align}
and in the particle density \cite{bruus2004many}
\begin{align}
  \label{eq:main63}
   n_{\rm p} &= - \frac {1/\pi}{L_x L_y}\int \d \epsilon \, n_{\rm F}(\epsilon) \, {\mb {Tr}}\, {\rm Im}\,G(\epsilon) ,
\end{align}
where $\hat v_x=\partial_{p_x}H_0$ is the velocity operator,
$G(\epsilon)$ is the full electron Green's function evaluated at $z=\epsilon+i0^+$, and the trace runs over all indices ${n,p_x,s}$.
Upon Poisson resummation, in both Eqs.\eqref{eq:main62},\eqref{eq:main63}, for a fixed $s$ the sum over level index $n$ is recast into a sum of harmonics at multiples of the frequency of the $s$ Fermi surface.
These combine with oscillations of the self-energy Eq.\eqref{eq:main61} so that $n_{\rm p}$ and $\sigma_{\rm L}$ contain terms at frequencies which are combinations of those of the $s$ and $\overline s$ bands.
In particular, the oscillating terms exhibited in Eq.\eqref{eq:main61} generate, to lowest order in powers of Dingle factors and $\Sigma/\mu$,
\emph{difference frequency}\cite{leeb2023theory} magneto-oscillations of $\sigma_{\rm L}$ and $n_{\rm p}$.
We denote by $ \tilde n_{\rm p}^{\rm df} $ and $ \tilde \sigma_{\rm L}^{\rm df}$ these leading-order contributions:
\begin{align}
   \label{eq:main73}
  \tilde \sigma_{\rm L}^{\rm df}&\approx \textstyle{\sum_{s,\lambda}}\, \ms A_{\sigma_{\rm L}}^{s,\lambda}(T)
                                  \, \cos \left ( 2\pi \xi_{s,\overline s}    \right )\\
                                &\times \left [ \sin \left ( 2\pi \bar \Omega_{\lambda} \frac{m_{\overline s}}{{\sf e}B} \right )
                                  - \sin \left ( 2\pi \bar \Omega_{\lambda} \frac{m_{s}}{{\sf e}B} \right )  \right ], \nonumber\\
  \label{eq:main76}
  \tilde n_{\rm p}^{\rm df} &\approx  \textstyle{\sum_{s,\lambda}} \, \ms A_{n_{\rm p}}^{s,\lambda}(T) \,
                              \sin \left ( 2\pi \xi_{s,\overline s} \right )  , 
\end{align}
with the shorthand $\xi_{s,\overline s} \equiv \left ( \frac{\mu_{s}m_s}{{\ms e}B} - \frac{\mu_{\overline s}m_{\overline s}}{{\ms e}B} \right )$,
and where $\ms A_{\sigma_{\rm L}}^{s,\lambda}(T)$ and $\ms A_{n_{\rm p}}^{s,\lambda}(T) $ are two amplitude factors,
explicit expressions of which are provided in the SM \cite{supp}.
 \begin{figure}[htbp]
  \begin{center}
    \includegraphics[width=.495 \columnwidth]{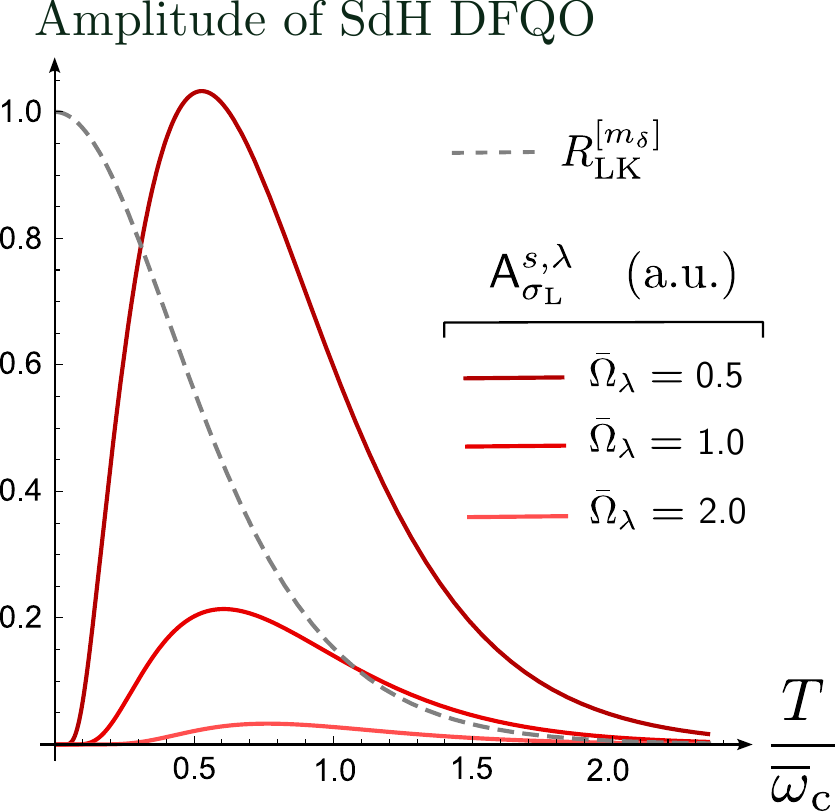} \hfill
      \includegraphics[width=.485 \columnwidth]{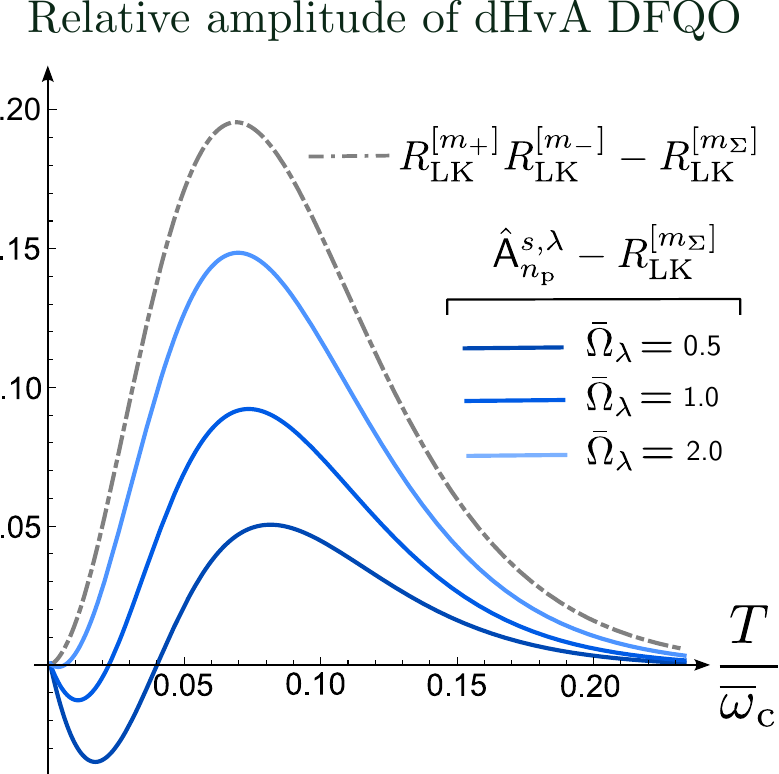}
      \caption{DFQO amplitudes as a function of temperature $T$ for several values of the boson frequency $\bar \Omega_\lambda$,
        with appropriate LK functions plotted in gray for comparison. Plots are identical for $s=+,-$.
        (left) SdH effect: $\ms A_{\sigma_{\rm L}}^{s,\lambda}(T)$ in arbitrary units.
        (right) dHvA effect: relative amplitude $\hat {\ms A}_{n_{\rm p}}^{s,\lambda} \equiv {\ms A}_{n_{\rm p}}^{s,\lambda}(T)/ {\ms A}_{n_{\rm p}}^{s,\lambda}(0)$ as a deviation from $R_{\rm LK}^{[m_\Sigma]}$ (see inset).}
\label{fig:plots}
\end{center}
\end{figure}
Their numerical evaluation, given in Fig.\ref{fig:plots},
was performed using the parameter values $m_+=0.9, m_-=1.1$,
and all energies and temperatures are in units of $\overline \omega_{\rm c}={\sf e}B/\overline m$ where $\overline m = \tfrac 1 2 (m_++m_-)$.

The dHvA effect Eq.\eqref{eq:main76} exhibits a unique difference frequency $|f_{s,\overline s}|$, identical for both $s=\pm$.
By contrast, the SdH effect Eq.\eqref{eq:main73} exhibits, for each $s=\pm$, two difference frequencies
$f_{s,\overline s} \pm \bar \Omega_{\lambda}m_{\overline s} /{\sf e}$ with the same amplitude.
For general boson dispersion this means that two symmetric pairs of peaks (or bumps)
on either side of the central difference frequency $f_{s,\overline s} $ appear in experiments. These four DFQO peaks can be better resolved for larger boson energies and a flatter dispersion around $\bar \Omega_{\lambda}$.

The temperature dependence of both the SdH and dHvA difference frequency effects, shown in Fig.\ref{fig:plots}, deviate strongly from the usual LK behavior.
Because conductivity, to the leading order in $g$, is driven by scattering with real (on-shell) bosons, 
that part of the fermion self-energy $\Sigma_{\rm int}$ responsible for SdH DFQO is thermally suppressed
at temperatures below the boson energy $\bar \Omega_\lambda$,
and the QO amplitude decays exponentially to zero at $1/T\rightarrow \infty$.
By contrast, thermodynamic quantities are also affected by the emission of virtual bosons, thus dHvA DFQO do not exhibit a thermal activation behavior. Strikingly, though, the thermal smearing of dHvA DFQO does not follow the mass-difference LK function $R_{\rm LK}^{[m_\delta]}(T)$ like in the SdH case, but instead follows roughly the mass-\emph{sum} LK function $R_{\rm LK}^{[m_\Sigma]}(T)$.

\textit{Discussion:}
We have shown how boson-mediated scattering leads to anomalous QO frequencies beyond the standard semi-classical picture. It is also well known that non-equilibrium effects can result in qualitatively new QO phenomena~\cite{dmitriev2012nonequilibrium} and  pumping the occupation of bosons can be used to tune quantum materials~\cite{de2021colloquium}. Here, we explore the potential effect of boson pumping 
 within a simple two temperature model~\cite{anisimov1974electron,caruso2022ultrafast}, i.e.\ assuming that the finite frequency pumped boson gas does not equilibrate with the system and remains at a temperature $T_b$ different than the entire crystal's temperature $T_0$, with the gas of fermions thermalizing at an intermediate value $T_f = \zeta T_b + (1-\zeta)T_0$, with $\zeta\in[0,1]$ a phenomenological parameter.
 \begin{figure}[htbp]
  \begin{center}
    \includegraphics[width=.485 \columnwidth]{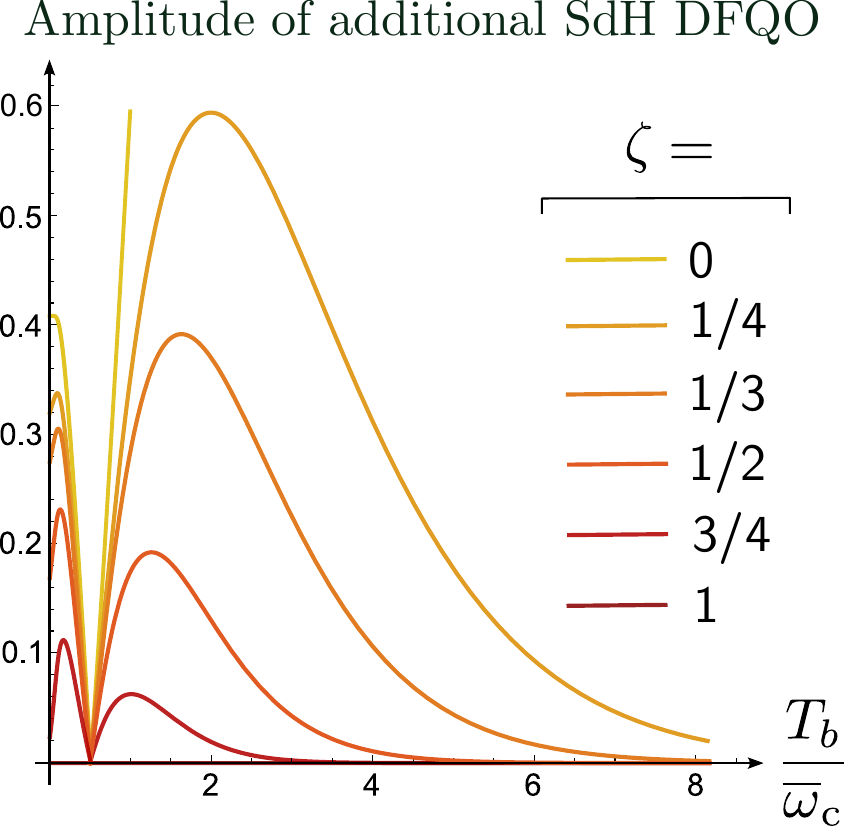} \hfill
      \includegraphics[width=.495 \columnwidth]{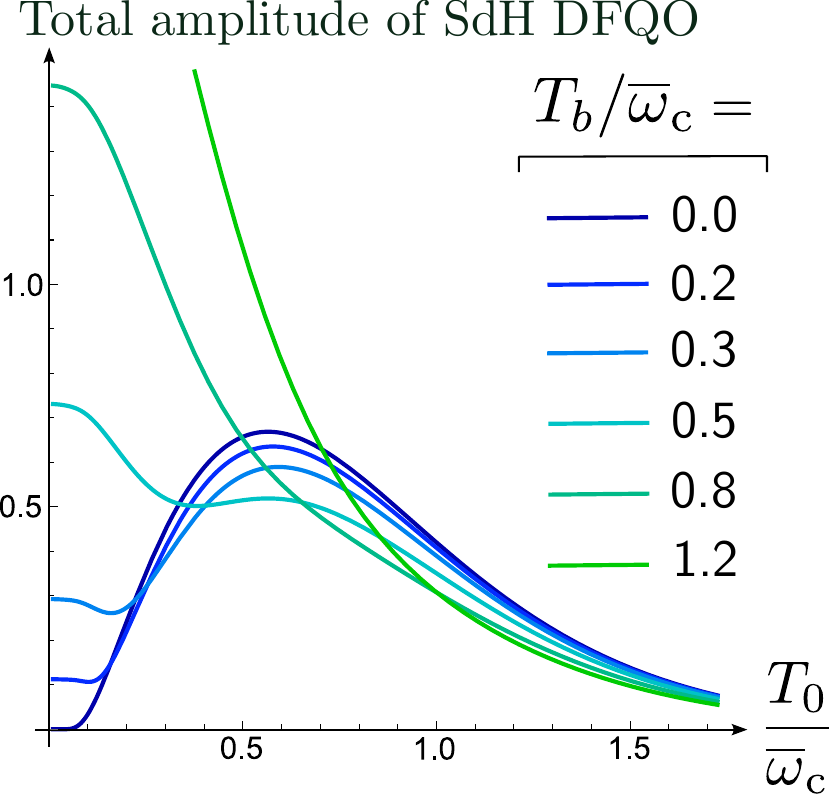}
\caption{(left) Amplitude $\big |{\sf A}_{\sigma_{\rm L},\rm 2T}^{s,\lambda} \big |$ of the additional SdH DFQO in the pumped boson case as a function of bosonic temperature $T_b$, for $T_0=0.5$ and several values of $\zeta$. (right) Total SdH DFQO amplitude $[({\sf A}_{\sigma_{\rm L}}^{s,\lambda}/2)^2+({\sf A}_{\sigma_{\rm L},\rm 2T}^{s,\lambda})^2]^{1/2}$ as a function of temperature $T_0$ for $\zeta=0$. (both) Units and parameter values are the same as in Fig.\ref{fig:plots}, and $\bar \Omega_\lambda=0.5$.}
\label{fig:plots-twoT}
\end{center}
\end{figure}
This non-equilibrium setup has no effect on the dHvA DFQO, but generates a new SdH DFQO contribution at the same frequency with $\pi/2$ dephasing and amplitude ${\sf A}_{\sigma_{\rm L},\rm 2T}^{s,\lambda}$, vanishing at $T_f=T_b$ (see Fig.\ref{fig:plots-twoT} and SM \cite{supp}).
Increasing $T_b$ has two competing effects: it increases the boson population $n_{\rm B}^{T_b}(\bar \Omega_\lambda)= 1/(e^{\bar \Omega_\lambda/T_b}-1)$, thereby enhancing the additional DFQO amplitude, and for $\zeta\neq 0$ it increases thermal smearing following $R_{\rm LK}^{[m_\delta]}(T_f)$. For fixed $T_b$, even at $\zeta=0$ the presence of a finite density of bosons suppresses the thermal activation behavior at $T_0\rightarrow 0$ and we, again, obtain a strong non-LK behavior. Interestingly, a controlled way of tuning the boson occupation in experiment would permit a determination of the electron boson coupling strength or extraction of the subsystems effective temperatures.

\textit{Conclusion:}
We have shown that interband scattering of electrons mediated by a dynamical boson generates difference frequency quantum oscillations (DFQO) which do not correspond to any semiclassical trajectory. More specifically, they exhibit a frequency splitting and anomalous temperature dependence, both of which are directly related to the boson dispersion $\Omega_{\lambda,\bs k}$, and their amplitude is proportional to the fermion-boson coupling strength $|g_{s,\overline s}^{\lambda,\bs k}|^2$. Thus boson-mediated DFQO carry information about the interacting fermion-boson system, and could be a useful experimental tool for probing electron interaction mechanisms in correlated metals. 

Our study poses a whole range of further questions, for example the possibility of such oscillations occurring in \emph{single-band} models on the verge of a density wave instability. There, the presence of a low-energy dynamical bosons could lead to precursor QO  prior to Fermi surface reconstruction. In addition, a microscopic treatment of non-equilibrium effects on DFQO will be essential for quantitative predictions. Finally, we have shown that long-overlooked QO phenomena appear from boson scattering within the Fermi liquid regime. Exploring the case of critical Fermi surfaces, where self-energy effects on QO are large~\cite{martin2003quantum,nosov2024entropy}, promises even more surprises. 

\vspace{1cm}

\acknowledgements

We thank Valentin Leeb for helpful conversations and comments on the manuscript.
L.M. thanks Pavel Nosov and J.K. thanks Achim Rosch for a helpful discussion.
We acknowledge support from the Imperial-TUM flagship
partnership, from the Deutsche Forschungsgemeinschaft 
(DFG, German Research Foundation) under Germany’s Excellence Strategy–EXC– 2111–390814868, DFG grants No. KN1254/1-2, KN1254/2-1, and TRR 360 - 492547816, as well as the Munich Quantum Valley, which is supported by the
Bavarian state government with funds from the Hightech Agenda Bayern Plus. This work was performed in part at Aspen Center for Physics, which is supported by National Science Foundation grant PHY-2210452.

\bibliography{Manuscript.bib}

\appendix

\begin{widetext}
  
  \section{Free theory in a magnetic field}
\label{sec:free-hamilt-eigensys}

We consider the free Hamiltonian:
\begin{align}
  \label{eq:1}
  H_0 &= \sum_{\bs p}\sum_{s =\pm} \left [ \frac 1 {2m_s} \left ( \alpha_s^2 (p_x-\text p_s^x-y)^2 +  {\alpha_s^{-2}}  (p_y-\text p_s^y)^2\right ) - \mu_s \right ]
      | \bs p,s \rangle \langle \bs p,s | .
\end{align}

\subsection{Diagonalization}
\label{sec:eigenfunctions}

Start from the standard diagonalization identity (where $p_y \equiv - i \partial_y$),
\begin{align}
  \label{eq:2}
\left [ - \left ( \partial_y \right )^2 + \left ( y - p_x \right )^2 -  \big ( 2n + 1 \big )  \right ]
   \psi_{n,p_x}(\bs r) &= 0,
\end{align}
with eigenfunctions $  \psi_{n,p_x}(\bs r) = \frac {1} {\sqrt {L_x}} {e}^{i p_x x} \psi_{n} \big (y - p_x \big ) $,
where $\psi_n(\xi) = e^{- \xi ^2 /2 } H_n \left [ \xi \right ] $ and $H_n$ is the $n$th Hermite polynomial
normalized such that $\int \text d \xi \,\psi_{n'}(\xi) \psi_n(\xi)=\delta_{n,n'}$. 
With the change of variables $y \rightarrow \alpha y$, it is straightforward to show that
\begin{align}
  \label{eq:7}
\left [  {\alpha^{-2}}\left ( -i \partial_y \right )^2 +  \alpha^2 \left ( y - p_x \right )^2 -  \big ( 2n + 1 \big )  \right ]
  \psi_{n, \alpha p_x}(x/\alpha, y \alpha) &= 0 ,
\end{align}
and using $ (- i \partial_y - \text p_y) \left ( e^{i  \text p_y y}  \;\cdot \,\right ) = e^{i  \text p_y y}   (-i \partial_y)$, 
\begin{align}
  \label{eq:8}
  \left [  {\alpha^{-2}}\left (-i \partial_y - \text p_y \right )^2 + \alpha^2  \left (y - (p_x-\text p_x) \right )^2
  - \big ( 2n + 1 \big ) \right ]
 e^{i  \text p_y y} \psi_{n, \alpha (p_x-\text p_x)}(x/\alpha, y \alpha)  &= 0.
\end{align}

Introducing orthogonal basis states $|s\rangle$ for the quantum number $s$,
this puts the free Hamiltonian into a diagonal form $H_0 = \sum_{s,n,p_x} \varepsilon_{s,n}  |s,n, p_x \rangle \langle s,n, p_x |$,
with energies $\varepsilon_{s,n} = \frac{{\ms e}B}{m_s} \left ( n+\tfrac 1 2\right )$,
and eigenstates $ |s,n, p_x \rangle = \psi_{s,n, p_x}  \otimes |s \rangle$, with eigenfunctions
\begin{align}
  \label{eq:27}
 \psi_{s,n, p_x} (\bs r) = \sqrt {\alpha_s/L_x}\, e^{i(p_x-\text p_s^x) x} e^{i  \text p_s^y y}
                                    \psi_{n} \big (\alpha_s(y -  (p_x- \text p_s^x )) \big )
\end{align}
normalized such that $ \langle s',n', p_x' | s,n, p_x \rangle=  \delta_{s,s'} \delta_{n,n'}\frac {2\pi}{L_x} \delta(p_x-p_x')$.

\subsection{Ladder operator}
\label{sec:ladder-operators}

Because of the recurrence properties
\begin{align}
  \label{eq:43}
   \partial_\xi \psi_n(\alpha \xi) &= \alpha \sqrt{n/2} \psi_{n-1}(\alpha\xi) - \alpha\sqrt{(n+1)/2} \psi_{n+1}(\alpha\xi) ,\\
 \alpha \xi \psi_n(\alpha \xi) &= \sqrt{n/2} \psi_{n-1}(\alpha \xi) + \sqrt{(n+1)/2} \psi_{n+1}(\alpha \xi) ,
\end{align}
the creation-destruction operators in the basis of Landau eigenstates of the band indexed by $s$ are
\begin{align}
  \label{eq:45}
  b_s^\dagger &=  \tfrac 1 {\sqrt 2} \left ( \alpha_s(p_x - y- \text p_s^x) + \alpha_s^{-1} i (p_y - \text p_s^y) \right ) ,\\
  b_s &= \tfrac 1 {\sqrt 2} \left (  \alpha_s(p_x - y- \text p_s^x) - \alpha_s^{-1} i (p_y - \text p_s^y)   \right ).
\end{align}
They verify
\begin{align}
  \label{eq:42ani}
  b_s^\dagger  (-1)^n \psi_{s,n,p_x} &= \sqrt{n+1} (-1)^{n+1} \psi_{s,n+1,p_x} ,\\
  b_s (-1)^n \psi_{s,n,p_x} &= \sqrt{n} (-1)^{n-1} \psi_{s,n-1,p_x} .
\end{align}

The velocity operator along axis $x$, defined as $\bs v_x = \partial_{\bs p_x}H_0$, reads in band $s$ 
\begin{align}
  \label{eq:22ani}
  v_s^x &= \alpha_s^2 (p_x - y - \text p_s^x)/m_s =  \alpha_s \frac{b_s + b_s^\dagger}{\sqrt 2m_s} .
\end{align}
One can also check that $H_0 = \sum_{s} \frac{{\ms e}B}{m_s} \left (  b_s^\dagger  b_s + \tfrac 1 2 \right )$.

\subsection{Free propagator}
\label{sec:free-propagators}

The non-interacting electron Green's function is
\begin{align}
  \label{eq:23}
  G_0(z) &=  \sum_{s=\pm}\sum_{n,p_x} \frac{ |s,n, p_x \rangle \langle s,n, p_x |} { z- \varepsilon_{s,n} } ,
\end{align}
with $z$ the complex frequency variable. Explicitly, written in components this reads
\begin{align}
  \label{eq:24}
   (G_0)_{s,s'}(\bs r, \bs r',z) 
  &= \delta_{s,s'}\sum_{n,p_x} \psi_{s,n,p_x}(\bs r) \left [  z - \varepsilon_{s,n} \right ]^{-1} \psi^*_{s,n,p_x}(\bs r').
\end{align}
Introducing $\bs R = \tfrac 1 2 (\bs r + \bs r')$, one can write after a few manipulations 
\begin{align}
  \label{eq:50}
  &\textstyle{\sum_{p_x} }\psi_{s,n,p_x}(\bs r) \psi^*_{s,n,p_x}(\bs r') \\
  &= e^{i\text p_s^y (y-y')} \int \frac{\text d^2 \bs k}{(2\pi)^2}  e^{i \bs k (\bs r - \bs r')}  2 \, (-1)^n
    \int \text d\xi  e^{-i k_y (2/ \alpha_s) (\xi - \alpha_s (R_y- k_x) ) } \psi_n(\xi - 2 \alpha_s( R_y -k_x) ) \psi_n(\xi) . \nonumber
\end{align}

We now apply the formula
\begin{align}
  \label{eq:26}
  \int \text d \xi e^{i \xi \lambda_x} \psi_n(\xi) \psi_n(\xi+\lambda_y)
  = L_n(\tfrac 1 2 \bs \lambda^2) e^{-\tfrac 1 4 \bs \lambda^2 } e^{-i \lambda_x \lambda_y/2}
\end{align}
where $L_n$ is the $n$th Laguerre polynomial, normalized such that $ \int_0^{\infty} \text dte^{-t} L_n(t) L_{n'}(t) = \delta_{n,n'}$,
to the particular case $\lambda_x = -(2/ \alpha_s) k_y$ and $\lambda_y=- 2 \alpha_s (R_y -k_x)$.
This yields
\begin{align}
  \label{eq:27}
 \textstyle{ \sum_{p_x}} \psi_{s,n,p_x}(\bs r) \psi^*_{s,n,p_x}(\bs r')
  &=  e^{i\text p_s^y (y-y')} \int \frac{\text d^2 \bs k}{(2\pi)^2} e^{i \bs k (\bs r - \bs r')}  e^{i (y+y') (x - x')/2}  (-1)^n 2
    L_n( 2 r_s^2(\bs k) ) e^{-r_s^2(\bs k)} ,
\end{align}
where we introduce the shorthand $r_s^2(\bs p)=  p_y^2/ \alpha_s^2 + p_x^2  \alpha_s^2 $. Thus,
\begin{align}
  \label{eq:14a}
  (G_0)_{s,s'} (\bs r, \bs r',z) &=  \delta_{s,s'} e^{i \varphi(\bs r,\bs r')} e^{i\text p_s^y (y-y')} 
                                           \int \frac{\text d^2 \bs p}{(2\pi)^2} e^{i \bs p (\bs r - \bs r')}  (\bar G_0)_{s,s} (\bs p,z) ,\\
 \label{eq:14b} \varphi(\bs r,\bs r') &:= (y+y')(x-x')/2,\\
 \label{eq:14c} (\bar G_0)_{s,s} (\bs p,z) &=   2 \sum_{n} (-1)^n 
    L_n( 2 r_s^2(\bs p) ) e^{- r_s^2(\bs p)} \left [  z - \epsilon_{s,n} \right ]^{-1} ,
\end{align}
and notably the translational dependence of $G_0(z)$ can be factorized as a mere phase factor.

\section{Interacting theory in a field}
\label{sec:inter-theory-field}

\subsection{Disorder scattering}
\label{sec:adding-disord-scatt}

We consider intraband local-in-space impurity scattering $ H_{\rm dis} = \sum_{s,\mb r} u(\mb r)  c^\dagger_{s,\mb r}  c^{\vphantom \dagger}_{s,\mb r} $,
where the disorder potential $u(\mb r)$ is a gaussian random variable with zero mean and two-point correlation
$\langle u(\mb r) u(\mb r') \rangle = \bar u_0^2 \delta(\mb r - \mb r')$.

The corresponding self-energy in real space, obtained by averaging over disorder realization, is
\begin{align}
  \label{eq:4}
  \Sigma^{\rm dis}_{s,s'}(\bs r,\bs r',z) &= \delta_{s,s'} \langle  u(\bs r) u(\bs r') \rangle G^{\rm dis}_{s,s'}(\bs r,\bs r',z)
                                            = \bar u_0^2 \delta_{s,s'} \delta(\bs r - \bs r') G^{\rm dis}_{s,s}(\bs r,\bs r,z)
\end{align}
where $G_{\rm dis}^{-1} = G_0^{-1}- \Sigma_{\rm dis}$ and we assumed the non-crossing approximation
and did not include any corrections from boson scattering.
We now assume, and will find self-consistently, a pure imaginary diagonal form of $ \Sigma_{\rm dis}$ in the basis of Landau eigenstates:
\begin{align}
  \label{eq:13}
   (\Sigma^{\rm dis})_{s,s'}^{n,n'}(p_x,p_x';z) 
  &= - i \Gamma_{s}^{\rm dis}\, {\rm sign}({\rm Im}(z))\, \delta_{s,s'} \delta_{n,n'} \frac{2\pi}{L_x}\delta(p_x-p_x') ,
\end{align}
where $\Gamma_{s}^{\rm dis}$ is to be determined. 
Expressing $G_{\rm dis}$ in the basis of Landau eigenstates and using the Poisson formula
\begin{align}
  \label{eq:3}
   \sum_{l=0}^\infty f(l) &= \sum_{k=-\infty}^{+\infty} \int_0^\infty \text d l \, e^{i2\pi k l} f(l),
\end{align}
to resum the Landau level summation, and using $\sum_{p_x}\psi_{s,n,p_x}(\bs r) \psi^*_{s,n,p_x}(\bs r) = \frac 1 {2\pi}$, one obtains
\begin{align}
  \label{eq:60}
  G^{\rm dis}_{s,s}(\bs r,\bs r,z) &\approx \frac{1}{2\pi} 
                                     \sum_{p \in \mathbb Z} (-1)^p \int_{1/2}^\infty \d u \frac{e^{i 2\pi p u}}
                                     {z+\mu_s + i \Gamma_{s}^{\rm dis}\, {\rm sign}({\rm Im}(z)) - \omega_{{\rm c},s}  u }.
\end{align}

Upon extending the lower bound of the integral to $-\infty$ and contour integration,
the $p$ contribution in the above sum comes with a $|p|$th power of the Dingle factor.
Thus we keep only $|p|\leq 1$, and absorb the real part of the $p=0$ component into a redefinition of $\mu_s$. This yields
\begin{align}
  \label{eq:10}
 G^{\rm dis}_{s,s}(\bs r,\bs r,z) & \approx - \frac i {2 \omega_{{\rm c},s}} \,{\rm sign}({\rm Im}(z))\,
                                 \left [ 1 -2 e^{-2\pi  \Gamma_{s}^{\rm dis}/ \omega_{{\rm c},s}} e^{i 2\pi \tfrac{z+\mu_s}{\omega_{{\rm c},s}}\, {\rm sign}({\rm Im}(z)) }\right ] .
\end{align}

The oscillating term in Eq.\eqref{eq:10} oscillates with the frequency of the $s$ band, therefore it cannot generate difference frequency quantum oscillations
to the leading order in Dingle factors, as we show later. Neglecting this contribution by keeping only the zeroth power of Dingle factors in Eq.\eqref{eq:10}, 
one finds the self-consistent value $\Gamma_{s}^{\rm dis}\equiv  \overline u_0^2 /2\omega_{{\rm c},s}$.
The fermionic Green's function dressed by disorder then reads
\begin{align}
  \label{eq:63}
    (G_{\rm dis})_{s,s'} (\bs r, \bs r',z) &=  \delta_{s,s'} e^{i \varphi(\bs r,\bs r')} e^{i\text p_s^y (y-y')} 
                                           \int \frac{\text d^2 \bs p}{(2\pi)^2} e^{i \bs p (\bs r - \bs r')}  (\bar G_{\rm dis})_{s,s} (\bs p,z) ,\\
 (\bar G_{\rm dis})_{s,s}  (\bs p,z) &=   2 \sum_{n} (-1)^n 
    L_n( 2 r_s^2(\bs p) ) e^{- r_s^2(\bs p)} \left [  z - \epsilon_{s,n} + i \Gamma_{s}^{\rm dis}\, {\rm sign}({\rm Im}(z))\right ]^{-1} .
\end{align}

\subsection{Electron-boson self-energy}
\label{sec:compute-self-energy-1}

We now consider the fermion-boson interaction hamiltonian $H_{\rm int}$ defined in the main text.
Because the boson changes the fermionic quantum number $s\leftrightarrow \overline s$, the fermionic self-energy is diagonal in $s$.
Because the boson is neutral, the fermionic self-energy factorizes in the same way as the free propagator,
and can be expressed as
\begin{align}
  \label{eq:89}
  \Sigma^{\rm int}_{s,s'}(\bs r,\bs r',z) &= e^{i\varphi(\bs r,\bs r')} e^{i\text p_s^y (y-y')} \int \frac{d^2 \bs p}{(2\pi)^2} e^{i \bs p (\bs r - \bs r')}
                                  \,\delta_{s,s'} \,\bar \Sigma^{\rm int}_{s,s}(\bs p,z).
 \end{align}
 To the order of one loop, using the bare bosonic propagator  $(D_0) _{\lambda,\lambda'} (\bs k,i \omega_{\rm m})
 = \delta_{\lambda,\lambda'} \frac{ 2 \Omega_{\lambda,\bs k}}{ (i \omega_{\rm m})^2 - \Omega_{\lambda,\bs k}^2 }$
and the dressed fermionic propagator $G_{\rm dis}$ for internal lines, one finds
 \begin{align}
   \label{eq:62}
   \bar \Sigma^{\rm int}_{s,s}(\bs p,z) &= -T \sum_{\omega_{\rm m}}  \sum_{\lambda} \int \frac{d^2 \bs k}{(2\pi)^2} \frac 1 {2 \Omega_{\lambda,\bs k}}
                                          \left | g_{s,\overline s,\bs p}^{\lambda,\bs k} \right |^2\,
                                          (D_0)_{\lambda,\lambda}(\bs k,i \omega_{\rm m}) \,\bar G^{\rm dis}_{\overline s,\overline s}(\bs p+\bs k,z + i\omega_{\rm m}),
 \end{align}
 where $\omega_{\rm m} = 2 \text m \pi T, \rm m \in \mathbb Z$.
 
The Matsubara summation yields, up to a correction which vanishes in the limit $\Gamma_{s}^{\rm dis}\rightarrow 0$,
\begin{align}
  \label{eq:93}
  \bar \Sigma^{\rm int}_{s,s'}(\bs p,z)  &\approx \delta_{s,s'}  \int \frac{\text d^2 \bs k}{(2\pi)^2}
                                           e^{- r_{\overline s}^2(\bs p+\bs k)}   \sum_{\lambda} \frac 1{ 2\Omega_{\lambda,\bs k} }
                                      \left | g_{s,\overline s}^{\lambda,\bs k} \right |^2 \nonumber \\
                                    & \times 2 \sum_{n}  (-1)^n  L_n \left ( 2 r_{\overline s}^2(\bs p+\bs k)\right )
                                      \left [  \frac{n_{\rm B}(\Omega_{\lambda,\bs k}) + n_{\rm  F}( \tilde \varepsilon_{\overline s,n}   )  }{z +\Omega_{\lambda,\bs k} - \tilde \varepsilon_{\overline s,n} }
    - ( \Omega_{\lambda,\bs k} \leftrightarrow - \Omega_{\lambda,\bs k} ) \right ] ,
\end{align}
with $\tilde \varepsilon_{\overline s,n} \equiv \varepsilon_{\overline s,n} - i \Gamma_{s}^{\rm dis}\, {\rm sign}({\rm Im}(z)) $.
The neglected contribution has poles at $z=i(2{\rm n}+1)\pi T -\Omega_{\lambda,\bs k}, {\rm n}\in \mathbb Z$, away from the real axis,
therefore it does not contain magneto-oscillations of $ \bar \Sigma^{\rm int}_{s,s'}(\bs p,z) $ and is thus not relevant in the present study.

Expressed in the basis of eigenfunctions of the free hamiltonian,
\begin{align}
  \label{eq:51}
 (\Sigma^{\rm int})_{s,s'}^{n,n'}(p_x,p_x';z) 
  & =  \int \frac{\text d^2 \bs q}{(2\pi)^2}\bar \Sigma^{\rm int}_{s,s} (\bs q,z) \times I_{s,s}^{n,n'}(p_x,p_x';\bs q)
\end{align}
where 
\begin{align}
  \label{eq:53}
  I_{s,s}^{n,n'}(p_x,p_x';\bs q)
  &\equiv  \int \text d^2 \bs r  \text d^2 \bs r' e^{i\varphi(\bs r,\bs r')} e^{i\text p_s^y (y-y')} e^{i \bs q (\bs r - \bs r')} \psi^*_{s,n,p_x}(\bs r) \psi_{s,n',p_x'}(\bs r') \nonumber \\
  &=  (-1)^n \; \frac {2\pi} {L_x} \delta(p_x-p_x') \,4\pi \int\text d \xi \,  e^{- i \xi 2 q_y /\alpha_s } \;
    \psi_n (\xi + \alpha_{s} q_x  ) \psi_{n'} (\xi- \alpha_{s} q_x ) \nonumber \\
  &= (-1)^n \; \frac {2\pi} {L_x} \delta(p_x-p_x') \,4\pi \left ( -2 \alpha_{s} q_x - i 2 q_y /\alpha_{s} \right )^{n'-n} \; L_n^{(n'-n)}(2 r_s^2(\bs q))\;e^{ -r_s^2(\bs q) },
\end{align}
where the second line was obtained after a few simple manipulations,
and the third line results from the identity
\begin{align}
  \label{eq:44}
   \int \text d \xi e^{i \xi \lambda_x} \psi_n(\xi) \psi_{n'}(\xi+\lambda_y)
  &= e^{-i \lambda_x \lambda_y/2} e^{-\tfrac 1 4 \bs \lambda^2 } \left ( \lambda_y + i \lambda_x \right )^{n'-n} L_n^{(n'-n)}(\tfrac 1 2 \bs \lambda^2) ,
\end{align}
applied to the specific case $\lambda_x= -2 q_y /\alpha_s$ and $\lambda_y= -2 \alpha_s q_x$,
where $L_n^{(m)}$ the $(n,m)$th associated Laguerre polynomial.  

Combined with Eq.\eqref{eq:93}, this yields Eq.\eqref{eq:main50} in the main text, where explicitly
\begin{align}
  \label{eq:59}
  I_{s,l}^{n,n'}(\bs k) &\equiv  2 \int \frac{\text d^2 \bs q}{(2\pi)^2} e^{ -r_s^2(\bs q) }  e^{- r_{\overline s}^2(\bs q+\bs k)}  \left ( -2 \alpha_{s}q_x - i 2 q_y /\alpha_{s} \right )^{n'-n}
                                     \;  (-1)^n  L_n^{(n'-n)}(2 r_s^2(\bs q)) \, (-1)^l  L_l \left ( 2 r_{\overline s}^2(\bs q+\bs k)\right ) .
\end{align}

  \subsection{Dressed fermionic Green's function}
\label{sec:dress-ferm-greens}

The full inverse Green's function $G^{-1}=G_{\rm dis}^{-1}- \Sigma^{\rm int}$, expressed in the basis of Landau eigenstates, is
\begin{align}
  \label{eq:64}
  (G^{-1})_{s,s'}^{n,n'}(p_x,p_x';z) &=  (G_0^{-1})_{s,s'}^{n,n'}(p_x,p_x';z) - (\Sigma^{\rm dis})_{s,s'}^{n,n'}(p_x,p_x';z) - (\Sigma^{\rm int})_{s,s'}^{n,n'}(p_x,p_x';z) .
\end{align}
It is diagonal in its $s$ and $p_x$ indices, as well as $p_x$-independent, therefore the Green's function has the form
\begin{align}
  \label{eq:65}
  G_{s,s'}^{n,n'}(p_x,p_x';z)
  &= \delta_{s,s'}\frac{2\pi}{L_x} \delta(p_x-p_x')\, \left ( \delta_{n,n'} ( z - \varepsilon_{s,n} )  - \Delta_{n,n'}^s (z) + i \Gamma_{n,n'}^s(z)  \right )^{-1}.
\end{align}

Besides, the off-diagonal in $(n,n')$ coefficients of $G^{-1}$ are always much smaller than its diagonal ones
provided that $\left | \Sigma^{\rm int}_{s,s} \right | \lesssim \Gamma^{\rm dis}_s$, which we will assume.
Now, for any matrix $\ms M_{n,n'}$ such that $\ms M_{n,n} = O(1)$ whereas $\ms M_{n\neq n'} = O(\delta)$ with $\delta$ a small parameter,
it is easy to show that $(\ms M^{-1})_{n,n} = 1/\ms M_{n,n} + O(\delta^2)$ -- for instance by using the general inversion formula
$\ms M^{-1}=\tilde {\ms M}^\top / \left |\ms M \right |$, with $\tilde {\ms M}$ the comatrix and $\left |\ms M \right | $ the determinant of $\ms M$.
Applying this to the particular case $\ms M = G^{-1}$, we obtain the diagonal elements
\begin{align}
  \label{eq:66}
  \Delta_{n,n}^s(z)  &\approx  {\rm Re}\, (\Sigma^{\rm int})_{s,s}^{n,n}(z),\\
   \Gamma_{n,n}^s(z)  &\approx  - {\rm Im}\,  (\Sigma^{\rm int})_{s,s}^{n,n}(z)  + \Gamma_{s}^{\rm dis}\,{\rm sign}({\rm Im}(z)) .
\end{align}

\section{Self-energy oscillations}
\label{sec:self-energy-oscill}

\subsection{General calculation}
\label{sec:gener-form-oscill}

The summation over level index $n$ in Eq.\eqref{eq:93} (relabeled $n\mapsto l$ here) can be performed using Poisson resummation.
This step relies on the fact that $ (-1)^l  L_l (x) $ can be approximated by a smooth function of $l$ a \emph{continuous} variable around $x \sim 4l+2$,
in terms of the Airy function.\cite{bateman1953higher}

We find 
\begin{align}
  \label{eq:73}
 \mc S_{\overline s}^\lambda(\bs q,\bs k;z)&\equiv  \sum_{l}  \frac{n_{\rm B}(\Omega_{\lambda,\bs k}) + n_{\rm  F}( \tilde \varepsilon_{\overline s,l}   )  }{z +\Omega_{\lambda,\bs k} - \tilde \varepsilon_{\overline s,l} }
    \, (-1)^l  L_l \left ( 2 r_{\overline s}^2(\bs q+\bs k)\right )\\
  &\approx \frac 1 {\omega_{{\rm c},\overline s}} \sum_{p\in\mathbb Z}\int_{1/2}^\infty \text dl \;e^{i 2\pi p l}  \,
    \frac{n_{\rm B}(\Omega_{\lambda,\bs k}) + n_{\rm  F}( \tilde \varepsilon_{\overline s,l}   )  }
    {l -\xi_{\lambda,\overline s}(\bs k, z) - i (\Gamma_{\overline s}^{\rm dis}/ \omega_{{\rm c},\overline s}){\rm sign}({\rm Im}(z)) }
    \, (-1)^l  L_l \left ( 2 r_{\overline s}^2(\bs q+\bs k)\right ) ,\nonumber
\end{align}
where $\xi_{\lambda,\overline s}(\bs k, z) = (z + \mu_{\overline s}+\Omega_{\lambda,\bs k}) /  \omega_{{\rm c},\overline s} $.

The $p=0$ term yields a non-oscillatory component in the self-energy, which we do not write explicitly in the following.
Extending the lower integration bound to $-\infty$ and performing the contour integration, one obtains
$ \mc S_{\overline s}^\lambda(\bs q,\bs k;z)= \mc S'_{\overline s,\lambda}(\bs q,\bs k;z)+ \mc S''_{\overline s,\lambda}(\bs q,\bs k;z)+\dots$,
where ``+ \dots'' contains the $p=0$ non-oscillatory term and $\mc S'$ (resp. $\mc S''$) results from zeros of the denominator (resp. poles of the numerator) in Eq.\eqref{eq:73}.
Explicitly, we find
\begin{align}
  \label{eq:14}
  \mc S'_{\overline s,\lambda}(\bs q,\bs k;z)&
                                              \approx (1/\omega_{{\rm c},\overline s})
                                                \; 2\pi i  {\rm sign}({\rm Im}(z)) \,\exp \big( 2\pi i \xi_{\lambda,\overline s}(z) {\rm sign}({\rm Im}(z)) \big )
                                                \,\exp ( -2\pi  \Gamma_{\overline s}^{\rm dis}/ \omega_{{\rm c},\overline s} ) \nonumber \\
   &\times  \left ( n_{\rm B}(\Omega_{\lambda,\bs k}) + n_{\rm  F}( z +\Omega_{\lambda,\bs k} )  \right ) \,
     \left [ (-1)^l  L_l \right ]_{l =\xi_{\lambda,\overline s}(\bs k, z)} \left ( 2 r_{\overline s}^2(\bs q+\bs k)\right ),\\
   \mc S''_{\overline s,\lambda}(\bs q,\bs k;z)&
                                                 \approx  - \frac 1 \pi \frac{2\pi^2 T /\omega_{{\rm c},\overline s}}{\sinh(2\pi^2T/\omega_{{\rm c},\overline s} )}
                                                 \frac{  \left [ (-1)^l  L_l \right ]_{l =\xi_{\overline s}(0)} \left ( 2 r_{\overline s}^2(\bs q+\bs k)\right )}
                                                 {z + \Omega_{\lambda,\bs k} + i\Gamma_{\overline s}^{\rm dis}{\rm sign}({\rm Im}(z)) } \, \sin (2\pi \xi_{\overline s}(0))   \label{eq:14new},
\end{align}
where in computing the Matsubara sum yielding Eq.\eqref{eq:14new} we used the first Matsubara frequency approximation,
which in the present case is equivalent to assuming a small Dingle factor -- see Appendix \ref{sec:diff-freq-observ} for details.
Note that $  \mc S''_{\overline s,\lambda}$ contains one fewer Dingle factor, and one more LK function, than $  \mc S'_{\overline s,\lambda}$.

Plugging Eqs.\eqref{eq:14},\eqref{eq:14new} into Eq.\eqref{eq:main50} in the main text, one obtains
\begin{align}
  \label{eq:76}
  \tilde \Sigma_{s,s'}^{n,n'}(p_x,p_x';z) &\approx \delta_{s,s'} \frac {2\pi} {L_x} \delta(p_x-p_x') \,
                                            2 \int \frac{\text d^2 \bs k}{(2\pi)^2}  \sum_{\lambda} \frac{2\pi}{\Omega_{\lambda,\bs k}}\,\left | g_{s,\overline s}^{\lambda,\bs k} \right |^2 \\
                                         &   \int \frac{\text d^2 \bs q}{(2\pi)^2} e^{ -r_s^2(\bs q) }  e^{- r_{\overline s}^2(\bs q+\bs k)}
                                            \left ( -2 \alpha_{s}q_x - i 2 q_y /\alpha_{s} \right )^{n'-n}    \;  (-1)^n  L_n^{(n'-n)}(2 r_s^2(\bs q)) \nonumber \\
   & \bigg [  (1/\omega_{{\rm c},\overline s}) \, R_{\rm D}^{\overline s} \, 2\pi i\, {\rm sign}({\rm Im}(z))
                                     \sum_{\eta=\pm} \eta \, \exp \left (2\pi i \,\xi_{\lambda,\overline s}^\eta(\bs k,z)\, {\rm sign}({\rm Im}(z)) \right )  \nonumber \\
   &\qquad  \left ( n_{\rm B}(\eta \Omega_{\lambda,\bs k}) + n_{\rm  F}( z +\eta \Omega_{\lambda,\bs k} )  \right )
     \left [ (-1)^l  L_l \right ]_{l =\xi_{\lambda,\overline s}^\eta(\bs k, z)} \left ( 2 r_{\overline s}^2(\bs q+\bs k)\right ) \nonumber \\
  &- \frac 1 \pi \frac{2\pi^2 T /\omega_{{\rm c},\overline s}}{\sinh(2\pi^2T/\omega_{{\rm c},\overline s} )}
   \, \sin (2\pi \xi_{\overline s}(0)) \,
    \sum_{\eta=\pm} \eta \, \frac { \left [ (-1)^l  L_l \right ]_{l =\xi_{\overline s}(0)} \left ( 2 r_{\overline s}^2(\bs q+\bs k)\right )}
    {z + \eta \Omega_{\lambda,\bs k} + i\Gamma_{\overline s}^{\rm dis}{\rm sign}({\rm Im}(z)) }  \bigg ],  \nonumber 
\end{align}
where the tilde in $ \tilde \Sigma $ symbolizes that the non-oscillatory part of the self-energy is not included in Eq.\eqref{eq:76},
and we introduced the shorthand $\xi_{\lambda,\overline s}^\eta(\bs k,z) = (z + \mu_{\overline s} + \eta \Omega_{\lambda,\bs k})/  \omega_{{\rm c},\overline s}$.

\subsection{Case of a single bosonic frequency}
\label{sec:case-single-bosonic}

We consider the case where all bosonic energies $\Omega_{\lambda,\bs k}$ in Eq.\eqref{eq:93} collapse to a single one $\bar \Omega_\lambda $.
This is exact when the bosonic band is flat, and more generally this is approximately the case
when most contributions to the integral come from values of the energy peaked around $\bar \Omega_\lambda  $.
  From now on we assume the replacement $\Omega_{\lambda,\bs k}\mapsto \bar \Omega_\lambda  $ can be performed in Eq.\eqref{eq:93}.
  
  Performing it in Eq.\eqref{eq:76}, one obtains
\begin{align}
  \label{eq:20}
  \tilde \Sigma_{s,s'}^{n,n'}(p_x,p_x';z) &\approx \delta_{s,s'} \frac {2\pi} {L_x} \delta(p_x-p_x')  \,\sum_{\lambda} (2\pi/ \bar \Omega_{\lambda}) \, \times \\
                                          & \bigg [ (1/\omega_{{\rm c},\overline s}) \, R_{\rm D}^{\overline s} \, 2\pi i\, {\rm sign}({\rm Im}(z))\,
                                            \sum_{\eta= \pm} \, {\rm exp}( 2\pi i \xi^\eta_{\lambda,\overline s}(z)  )
                                            \left [ n_{\rm B}(\bar \Omega_{\lambda}) + n_{\rm  F}( \eta z + \bar \Omega_{\lambda} ) \right ]
                                            \,\times \mc I_{n,n'}^{\lambda;s;\eta}(z) \nonumber \\
  & - \frac 1 \pi \frac{2\pi^2 T /\omega_{{\rm c},\overline s}}{\sinh(2\pi^2T/\omega_{{\rm c},\overline s} )}
   \, \sin (2\pi \xi_{\overline s}(0)) \,
    \sum_{\eta=\pm} \eta \, \frac { 1 }
    {z + \eta \bar{\Omega}_\lambda + i\Gamma_{\overline s}^{\rm dis}{\rm sign}({\rm Im}(z)) }   \,\times \mc I_{n,n'}^{\lambda;s;0}(0) \bigg ] \nonumber ,
\end{align}
where
\begin{align}
  \label{eq:8}
  \mc I_{n,n'}^{\lambda;s;\eta}(z) &\equiv 2\int \frac{\text d^2 \bs q}{(2\pi)^2} e^{ -r_s^2(\bs q) }
                                     (-1)^n  L_n^{(n'-n)}(2 r_s^2(\bs q)) \left ( -2 a_{s} q_x - i 2 q_y /a_{s} \right )^{n'-n} \nonumber \\
                                   & \times \int \frac{\text d^2 \bs k}{(2\pi)^2} \left | g_{s,\overline s}^{\lambda,\bs k} \right |^2 e^{- r_{\overline s}^2(\bs q+\bs k)}
                                     \left [ (-1)^l  L_l \right ]_{l = \xi^\eta_{\lambda,\overline s}(z) }\left ( 2 r_{\overline s}^2(\bs q+\bs k)\right ) 
\end{align}
and one used the shorthand $\xi_{\lambda,\overline s}^\eta(z) = (z + \mu_{\overline s} + \eta \bar \Omega_{\lambda})/  \omega_{{\rm c},\overline s}$
 as well as the identity $ \eta \left [ n_{\rm B}(\eta\bar \Omega_{\lambda}) + n_{\rm  F}( z + \eta \bar \Omega_{\lambda} ) \right ]
  = \left [ n_{\rm B}(\bar \Omega_{\lambda}) + n_{\rm  F}( \eta z + \bar \Omega_{\lambda} ) \right ]$ -- the latter holds even when fermions and bosons are at different temperatures.

  In the particular case $n=n'=\xi_s(\epsilon)$ and $z=\epsilon+i0^+$ this yields Eq.\eqref{eq:main61} in the main text,
  where $\delta_{s,s'} \frac {2\pi} {L_x} \delta(p_x-p_x') $ is not written to avoid clutter, oscillating terms are proportional to 
  \begin{align}
    \label{eq:77a}
   i ( {\sf \Sigma}')_{s,\overline s}^{\lambda,\eta}(\epsilon)
    &= (1/\omega_{{\rm c},\overline s})  \, 2\pi i \, R_{\rm D}^{\overline s} \, (2\pi/ \bar \Omega_{\lambda})
      \, \left [ n_{\rm B}(\bar \Omega_{\lambda}) + n_{\rm  F}( \eta \epsilon + \bar \Omega_{\lambda} ) \right ]  \, \mc I_{\xi_s(\epsilon), \xi_s(\epsilon)}^{\lambda;s;\eta}(\epsilon), \\
      \label{eq:77b}
    ( {\sf \Sigma}'')_{s,\overline s}^{\lambda,\eta}(\epsilon)
    &= - \frac 1 \pi \, R_{\rm LK}^{[m_{\overline s}]}(T) \, (2\pi/ \bar \Omega_{\lambda})\,
   \frac { \eta }   {\epsilon + \eta \bar{\Omega}_\lambda + i\Gamma_{\overline s}^{\rm dis} }   \, \mc I_{\xi_s(\epsilon), \xi_s(\epsilon)}^{\lambda;s;0}(0),
  \end{align}
and $ {\sf \Sigma}_s(\epsilon) $ is the non-oscillatory part originating from the $p=0$ term in Eq.\eqref{eq:73} --
which is not relevant to our study of composite frequency oscillations.

One can note that $\mc I_{\xi_s(\epsilon), \xi_s(\epsilon)}^{\lambda;s;\eta\rightarrow 0}(0) \in \mathbb R$,
therefore when this replaces $\mc I$ in Eq.\eqref{eq:77a}, $( {\sf \Sigma}')_{s,\overline s}^{\lambda,\eta}(\epsilon)$ is purely real.
Meanwhile $ ( {\sf \Sigma}'')_{s,\overline s}^{\lambda,\eta}(\epsilon)$ is complex because of $ i\Gamma_{\overline s}^{\rm dis} $ in the denominator.

\section{CFQO in observables}
\label{sec:diff-freq-observ}

\subsection{Shubnikov-de Haas effect}
\label{sec:shubnikov-de-haas}

\subsubsection{General calculations}
\label{sec:general-calculations-1}

We compute $\sigma_{xx}$ at finite temperature using the convolution formula
\begin{align}
  \label{eq:67a}
  \sigma_{xx}&= \int \text d \epsilon \frac{\hat \sigma_{xx}(\epsilon)}{4T \cosh^2 \left ( \frac{\epsilon }{2T} \right )},\\
    \label{eq:67b}
\hat \sigma_{xx}(\epsilon) &= -\frac{\ms e^2}{\pi } \frac 1 {L_x L_y}\,\mb {Tr}_{n,p_x,s}\left [ \hat v_x \,{\rm Im} G(\epsilon) \,\hat v_x \,{\rm Im} G(\epsilon) \right ].
\end{align}
Clearly, matrix elements of $G$ with $n\neq n'$ can only contribute in Eq.\eqref{eq:67b} if they appear twice.
Because they are much smaller than diagonal elements, we keep only contribution from diagonal elements $G_{n,n}(\epsilon)$, which yields

\begin{align}
  \label{eq:9a}
   \hat \sigma_{xx}(\epsilon)
   &=- \frac{\ms e^2}{2\pi^2} \sum_{s=\pm} \frac {\alpha_s^2} {m_s^2}
     \sum_{n=0}^\infty  (n+1) \,{\rm Im} G_{s,s}^{n,n} (\epsilon)  \, {\rm Im}  G_{s,s}^{n+1,n+1} (\epsilon) ,\\
       \label{eq:9b}
      {\rm Im} G_{s,s}^{n,n} (\epsilon) &= -  \Gamma_n^s(\epsilon) /
\left \{ \left [ \epsilon - \varepsilon_{s,n} - \Delta_n^s(\epsilon) \right ]^2 + \Gamma_n^s(\epsilon)^2 \right \} ,
\end{align}
where Eq.\eqref{eq:9b} derives from Eq.\eqref{eq:65}.

We perform Poisson-resummation of the sum over $n$ in Eq.\eqref{eq:9a} into a sum over $p \in \mathbb Z$ the harmonics index.
We extend the lower integration bound to $-\infty$, then perform complex contour integration.

Keeping only terms to first order in Dingle factors ($|p|\leq 1$), and focusing on the oscillatory contributions ($p \neq 0$)
denoted by a tilde, we find  
\begin{align}
  \label{eq:81}
  \tilde \sigma_{xx}(\epsilon)
  &\approx \frac{ \ms e^2}{\pi} \sum_{s=\pm} \alpha_s^2 \frac { \Gamma^s_{\xi_s(\epsilon)}(\epsilon) /  \omega_{{\rm c},s}  }
    {  1 + 4 \left ( \Gamma^s_{\xi_s(\epsilon)}(\epsilon) /  \omega_{{\rm c},s}  \right )^2 } \,
    \left (  \xi^\star_{s}(\epsilon) e^{i 2\pi \xi^\star_{s}(\epsilon) }  +   {\rm c.c.} \right ) ,
\end{align}
where $\xi^\star_{s}(\epsilon) \equiv \xi_{s}(\epsilon) - ( \Delta_{\xi_s(\epsilon)}^s (\epsilon)  - i \Gamma_{\xi_s (\epsilon)}^s (\epsilon) )/\omega_{{\rm c},s}$
and we recall the definition $\xi_s(\epsilon) = (\epsilon+\mu_{s})/\omega_{{\rm c},s} $.

In Eq.\eqref{eq:81}, composite frequency oscillations arise, to the lowest order in $\Sigma_{\rm int}/\mu \ll 1$,
from keeping the (dominant) non-oscillatory part of $\xi^\star_{s}(\epsilon) $ and that part of the ratio-like prefactor
oscillating with the frequency of the $\overline s$ band, namely 
\begin{align}
  \label{eq:79}
  \alpha_s^2 \frac { \Gamma^s_{\xi_s(\epsilon)}(\epsilon) /  \omega_{{\rm c},s}  }
  {  1 + 4 \left ( \Gamma^s_{\xi_s(\epsilon)}(\epsilon) /  \omega_{{\rm c},s}  \right )^2 }
  &\approx \sum_\lambda\sum_{\eta=\pm} \left \{  (\breve {\sf \Sigma}')_{s,\overline s}^{\lambda,\eta}(\epsilon)
    \,   \cos \left ( 2\pi \xi_{\lambda,\overline s}^\eta (\epsilon) \right )
    + (\breve {\sf \Sigma}'')_{s,\overline s}^{\lambda,\eta}(\epsilon)
    \,   \sin \left ( 2\pi \xi_{\overline s}(0) \right )   \right \} \; + \dots ,
\end{align}
where we recall $\xi_{\lambda,\overline s}^\eta(z) = (z + \mu_{\overline s} + \eta \bar \Omega_{\lambda})/  \omega_{{\rm c},\overline s}$.
Here ``+ \dots'' contains the non-oscillatory contribution, the contribution (arising from disorder scattering) which oscillates at the frequency of the $s$ band,
as well as terms with higher powers of the Dingle factor and of $\Sigma_{\rm int}/\mu \ll 1$.

To the first order in $\left | {\rm Im}(\Sigma^{\rm int}_{s,s}) \right | / \Gamma^{\rm dis}_s \lesssim 1$, the oscillation amplitudes are
\begin{align}
  \label{eq:9}
(\breve {\sf \Sigma}')_{s,\overline s}^{\lambda,\eta}(\epsilon) =
 \frac{ \alpha_s^2}{\omega_{{\rm c},s}} \,\frac{4 (\Gamma_{s}^{ \rm dis}/\omega_{{\rm c},s})^2 -1}{\left ( 4 (\Gamma_{s}^{ \rm dis}/\omega_{{\rm c},s})^2 + 1 \right )^2 } \,( {\sf \Sigma}')_{s,\overline s}^{\lambda,\eta}(\epsilon),
 \quad (\breve {\sf \Sigma}'')_{s,\overline s}^{\lambda,\eta}(\epsilon) =
 \frac{ \alpha_s^2}{\omega_{{\rm c},s}}  \,\frac{4 (\Gamma_{s}^{ \rm dis}/\omega_{{\rm c},s})^2 -1}{\left ( 4 (\Gamma_{s}^{ \rm dis}/\omega_{{\rm c},s})^2 + 1 \right )^2 } \,{\rm Im}( {\sf \Sigma}'')_{s,\overline s}^{\lambda,\eta}(\epsilon).
\end{align}
A higher-order expansion modifies the specific value of $ \breve {\sf \Sigma}_{s,\overline s}^{\lambda,\eta}(\epsilon)$ but not the general form of the result Eq.\eqref{eq:79}.
In Eqs.\eqref{eq:79},\eqref{eq:9} we used that, to the order of the approximations we make in the following, $( {\sf \Sigma}')_{s,\overline s}^{\lambda,\eta}(\epsilon) \in \mathbb R$.

\subsubsection{Main DFQO contribution}
\label{sec:main-dfqo-contr}

To the lowest order in $\Sigma_{\rm int}/\mu \ll 1$,
approximating $\xi^\star_{s}(\epsilon) \approx \xi_{s}(\epsilon) - i \Gamma_s^{\rm dis}$ in the exponential factors, 
\begin{align}
  \label{eq:84a}
  \tilde \sigma^{\rm df}_{xx}(\epsilon) &\approx \frac{{\ms e}^2}{\pi} \sum_{s,\lambda}\sum_{\eta=\pm} (\breve {\sf \Sigma}')_{s,\overline s}^{\lambda,\eta}(\epsilon)
                                 \, \cos \left ( 2\pi (\xi_{\lambda,\overline s}^\eta(\epsilon) - \xi_{s} (\epsilon) )\right )
                                 \, \xi_s (\epsilon) \,  e^{ -2\pi  \Gamma_{s}^{\rm dis}/ \omega_{{\rm c}, s} } \\
   \label{eq:84b}
& - \frac{{\ms e}^2}{\pi} \sum_{s,\lambda}\sum_{\eta=\pm} (\breve {\sf \Sigma}'')_{s,\overline s}^{\lambda,\eta}(\epsilon)
                                 \, \sin \Big ( 2\pi (\xi_{s} (\epsilon) - \xi_{\overline s}(0) ) \Big )
                                 \, \xi_s (\epsilon) \,  e^{ -2\pi  \Gamma_{s}^{\rm dis}/ \omega_{{\rm c}, s} } + \dots ,
\end{align}
where ``+\dots'' contains all terms which are either not difference frequency or higher order in Dingle factors or $\Sigma_{\rm int}/\mu$.
Eq.\eqref{eq:84a} comes with two Dingle factors since $(\breve {\sf \Sigma}')_{s,\overline s}^{\lambda,\eta}(\epsilon) \propto  R_{\rm D}^{\overline s} $.
Meanwhile Eq.\eqref{eq:84b} comes with only one Dingle factor, however it is strongly decreasing with temperature
since $(\breve {\sf \Sigma}'')_{s,\overline s}^{\lambda,\eta}(\epsilon) \propto  R_{\rm LK}^{[m_{\overline s}]} (T)$.
Therefore at large temperatures the leading difference frequency contributions arise solely from Eq.\eqref{eq:84a}
and we temporarily neglect Eq.\eqref{eq:84b} in the following -- we will show later that within our approximations, it actually vanishes.

We now compute the energy integral Eq.\eqref{eq:67a}.
In Eq.\eqref{eq:84a} we approximate the outer $\xi_s (\epsilon) \rightarrow \xi_s (0)$,
and in Eq.\eqref{eq:77a} we take $\mc I_{\xi_s(\epsilon), \xi_s(\epsilon)}^{\lambda;s;\eta}(\epsilon) \rightarrow \mc I_{\xi_s(0), \xi_s(0)}^{\lambda;s;\eta\rightarrow 0}(0)$.
Thus $ (\breve {\sf \Sigma}')_{s,\overline s}^{\lambda,\eta}(\epsilon)
= \left [ n_{\rm B}(\bar \Omega_{\lambda}) + n_{\rm  F}( \eta \epsilon + \bar \Omega_{\lambda} ) \right ] \, (\check {\sf \Sigma}')_{s,\overline s}^{\lambda} \in \mathbb R$.

The leading difference frequency oscillatory term in electrical conductivity thus reads
\begin{align}
  \label{eq:57}
  \tilde \sigma_{xx}^{\rm df}
  &\approx \frac{{\ms e}^2}{\pi}   {\rm Re}  \sum_{s,\lambda}\sum_{\eta=\pm}\,\xi_s (0) \, 
    e^{ i 2\pi \left [ \xi_{\lambda,\overline s}^\eta(0) - \xi_{s}(0) \right ] } \, R_{\rm D}^{s}\, 
    (\check {\sf \Sigma}')_{s,\overline s}^{\lambda}
    \, \int \text d \epsilon \,  \frac{e^{ i 2\pi \left[  m_{\overline s} - m_s\right ] \epsilon }}{4T \cosh^2 \left ( \frac{\epsilon }{2T} \right )} \,
    \left [ n_{\rm B}(\bar \Omega_{\lambda}) + n_{\rm  F}( \eta \epsilon + \bar \Omega_{\lambda} ) \right ] ,\\
  (\check {\sf \Sigma}')_{s,\overline s}^{\lambda}
  &\equiv
\; 2 \frac{1}{\omega_{{\rm c},\overline s}} \, 2\pi \, e^{ -2\pi  \Gamma_{\overline s}^{\rm dis}/ \omega_{{\rm c},\overline s} } \, \frac{2\pi}{\bar \Omega_{\lambda}}
    \, \mc I_{\xi_s(0), \xi_s(0)}^{\lambda;s;\eta\rightarrow 0}(0)\, \frac{\alpha_s^2}{\omega_{{\rm c},s}}
    \frac{4 (\Gamma_{s}^{ \rm dis}/\omega_{{\rm c},s})^2 -1}{\left ( 4 (\Gamma_{s}^{ \rm dis}/\omega_{{\rm c},s})^2 + 1 \right )^2 } .
\end{align}

Using the identities (for $\eta=\pm$)
\begin{align}
  \label{eq:25a}
   \int \text d \epsilon \frac{e^{ i 2\pi X \epsilon }}{4T \cosh^2 \left ( \frac{\epsilon }{2T} \right )} 
  &= \frac{2\pi^2 X T}{\sinh(2\pi^2 X T)} ,\\
  \int \text d \epsilon \frac{e^{ i 2\pi X \epsilon }}{4T \cosh^2 \left ( \frac{\epsilon }{2T} \right )} n_{\rm  F}( \eta \epsilon + \bar \Omega_{\lambda} ) 
 \label{eq:25b} &= -\frac{2\pi^2 X T}{\sinh(2\pi^2 X T)} n_{\rm B}(\bar \Omega_\lambda) - \frac{2\pi T \sin(\pi X  \bar \Omega_\lambda )}{\sinh(2\pi^2 X T)}\,
  n'_{\rm B}(\bar \Omega_\lambda) \,  e^{-i\pi \eta X \bar \Omega_\lambda},
\end{align}
where $n'_{\rm B} =\partial_\epsilon n_{\rm B}$, and computing the sum over $\eta=\pm$, one obtains Eq.\eqref{eq:main73} in the main text, where
\begin{align}
  \label{eq:18}
 \ms A_{\sigma_{\rm L}}^{s,\lambda}(T)   &=\frac{{\ms e}^2}{\pi}  \,\xi_s (0) \,
                                      \, R_{\rm D}^{s}\; (\check {\sf \Sigma}')_{s,\overline s}^{\lambda}  \,\left (-n'_{\rm B}(\bar \Omega_\lambda) \right )\,
                                      \frac{2\pi T }{\sinh(2\pi^2 \left[  m_{\overline s} - m_s\right ]   T)} .
\end{align}

When $T_b\neq T_f$, a new contribution appears at the same difference frequencies,
from the absence of cancellation in $n_{\rm B}(\bar \Omega_\lambda) \text{Eq.\eqref{eq:25a}} + \text{Eq.\eqref{eq:25b}}$:
\begin{align}
  \label{eq:82a}
\tilde \sigma_{xx}^{\rm df,2T} &\approx \sum_{s,\lambda} \ms A_{\sigma_{\rm L},\rm 2T}^{s,\lambda}(T_b,T)  \sum_{\eta=\pm}\,  
                                 \cos \left (  2\pi ( \xi_{\lambda,\overline s}^\eta(0) - \xi_{s}(0) ) \right ) ,\\
  \label{eq:82b}
  \ms A_{\sigma_{\rm L},\rm 2T}^{s,\lambda}(T_b,T) &\equiv   \frac{{\ms e}^2}{\pi}\,\xi_s (0)
                                                     \, R_{\rm D}^{s}\, (\check {\sf \Sigma}')_{s,\overline s}^{\lambda} \;
                                                   \frac{2\pi^2  \,  \left[  m_{\overline s} - m_s\right ]  T}{\sinh(2\pi^2  \left[  m_{\overline s} - m_s\right ]  T)} \,
  \left ( n^{T_b}_{\rm B}(\bar \Omega_{\lambda}) - n^{T}_{\rm B}(\bar \Omega_{\lambda}) \right ).
\end{align}

\subsubsection{Other difference and sum frequency contributions}
\label{sec:other-difference-sum}

We now address the other difference frequency contribution arising from Eq.\eqref{eq:84b}.
We take $\xi_s (\epsilon) \rightarrow \xi_s (0)$, so that Eq.\eqref{eq:77b} yields $\sum_{\eta=\pm} ({\sf \Sigma}'')_{s,\overline s}^{\lambda,\eta}(\epsilon)
= (\hat {\sf \Sigma}'')_{s,\overline s}^{\lambda}\sum_{\eta=\pm}\eta(\epsilon+\eta \bar \Omega_\lambda + i \Gamma_{\overline s}^{\rm dis} )^{-1}$,
and the extra difference frequency contribution reads
\begin{align}
  \label{eq:17}
  \tilde \sigma_{xx}^{\rm df,other} &= - \frac{{\ms e}^2}{\pi}    \sum_{s,\lambda} \xi_s (0) \, R_{\rm D}^s \,
                                      (\hat {\sf \Sigma}'')_{s,\overline s}^{\lambda} \;
                                    {\rm Im}  \int \text d \epsilon \frac{\sin \Big ( 2\pi (\xi_{s} (\epsilon) - \xi_{\overline s}(0) ) \Big ) }{4T \cosh^2 \left ( \frac{\epsilon }{2T} \right )}\,
                                   \sum_{\eta=\pm}   \frac {\eta} {\epsilon+\eta \bar \Omega_\lambda + i \Gamma_{\overline s}^{\rm dis} }.
\end{align}
The following integral
\begin{align}
  \label{eq:21}
   \int \text d \epsilon \frac{e^{i 2\pi m_s \epsilon }}{4T \cosh^2 \left ( \frac{\epsilon }{2T} \right )}\,
  \sum_{\eta=\pm}   \frac {\eta} {\epsilon+\eta \bar \Omega_\lambda + i \Gamma_{\overline s}^{\rm dis} }
  &= - i 2\pi T \sum_{\rm n \geq 0}e^{-(2{\rm n}+1)2\pi^2 m_s T} \\
  &\times \sum_{\eta=\pm}  \eta
    \left ( \tfrac{2\pi m_s}{ \Gamma_{\overline s}^{\rm dis} + (2{\rm n}+1)\pi T - i \eta \bar \Omega_\lambda }
    +   \tfrac{1}{\left ( \Gamma_{\overline s}^{\rm dis} + (2{\rm n}+1)\pi T - i \eta \bar \Omega_\lambda  \right )^2 }  \right ) \nonumber
\end{align}
is real, therefore the energy integral in Eq.\eqref{eq:17} is real and $\tilde \sigma_{xx}^{\rm df,other} =0$.
Thus the only difference frequency contribution is that of Eq.\eqref{eq:57}.

For the sake of completeness, we now also address the sum frequency contributions.
Similarly to the above, there is no contribution from $  (\hat {\sf \Sigma}'')_{s,\overline s}^{\lambda}$, and the only sum frequency term is
obtained by replacing $- \xi_{s} (\epsilon)\rightarrow + \xi_{s} (\epsilon)$ in Eq.\eqref{eq:84a}, which yields
\begin{align}
  \label{eq:22}
   \tilde \sigma_{xx}^{\rm sf}
  &\approx  \sum_{s,\lambda} \ms A_{\sigma_{\rm L},{\rm sf}}^{s,\lambda}(T)
    \, \cos \left ( 2\pi (\xi_{s}(0) + \xi_{\overline s} (0))   \right ) \, \left [ \sin \left ( 2\pi \bar \Omega_{\lambda} \frac{m_{\overline s}}{{\sf e}B} \right )
                                  + \sin \left ( 2\pi \bar \Omega_{\lambda} \frac{m_{s}}{{\sf e}B} \right )  \right ],\\
  \ms A_{\sigma_{\rm L},{\rm sf}}^{s,\lambda}(T)
  &\equiv \frac{{\ms e}^2}{\pi}  \,\xi_s (0) \,
                                      \, R_{\rm D}^{s}\; (\check {\sf \Sigma}')_{s,\overline s}^{\lambda}  \,\left (-n'_{\rm B}(\bar \Omega_\lambda) \right )\,
                                      \frac{2\pi T }{\sinh(2 \pi^2 (m_s+m_{\overline s}) T)}.
\end{align}

When $T_b\neq T_f$, an extra contribution is also generated, following the very same line of reasoning as that leading to Eqs.\eqref{eq:82a},\eqref{eq:82b}.
We do not write it explicitly here.

\subsection{De Haas-van Alphen effect}
\label{sec:de-haas-van}

\subsubsection{General calculations}
\label{sec:general-calculations}

We compute $n_{\rm p}$ at finite temperature using the convolution formula
\begin{align}
  \label{eq:7}
  n_{\rm p} &= \frac{1}{L_x L_y}\int_{-\infty}^{+\infty}\d \epsilon \, n_{\rm F}(\epsilon) \, \rho(\epsilon) ,\\
\rho(\epsilon) &= - \frac 1 \pi {\mb {Tr}}\, {\rm Im}\,G(\epsilon) = - \frac 1 \pi  {\rm Im}\sum_{s=\pm}\sum_{n=0}^\infty  G_{s,s}^{n,n}(\epsilon).
\end{align}

We perform Poisson-resummation of the sum over $n$ in Eq.\eqref{eq:7} into a sum over $p \in \mathbb Z$ the harmonics index.
We extend the lower integration bound to $-\infty$, then perform complex contour integration.
Keeping only terms to first order in Dingle factors ($|p|\leq 1$), and focusing on the oscillatory contributions ($p \neq 0$)
denoted by a tilde, we find  
\begin{align}
  \label{eq:19}
  \tilde \rho(\epsilon)   &\approx  - N_\phi\sum_{s=\pm} \frac 1 { \omega_{{\rm c},s} } \left ( e^{i 2\pi \xi^\star_{s}(\epsilon)} +\rm h.c. \right ) ,
\end{align}
where we recall $\xi^\star_{s}(\epsilon) \equiv \xi_{s}(\epsilon) - ( \Delta_{\xi_s(\epsilon)}^s (\epsilon)  - i \Gamma_{\xi_s (\epsilon)}^s (\epsilon) )/\omega_{{\rm c},s}$.
Composite frequency oscillations arise from that part of $\xi^\star_{s}(\epsilon) $ oscillating with the frequency of the $\overline s$ band, namely
\begin{align}
  \label{eq:29}
  \xi^\star_{s}(\epsilon) &\approx  \xi_{s}(\epsilon)   + i \Gamma_s^{\rm dis}/\omega_{{\rm c},s} + \dots \\
  &- \frac 1 { \omega_{{\rm c},s} } \sum_\lambda \sum_{\eta=\pm} i ({\sf \Sigma}')_{s,\overline s}^{\lambda,\eta}(\epsilon)
                             \,    e^{i 2\pi \left [ \xi_{\overline s}(\epsilon) + \eta \frac {\bar{\Omega}_\lambda}{\omega_{{\rm c},\overline s}} \right ] }
         - \frac 1 { \omega_{{\rm c},s} } \sum_\lambda \sum_{\eta=\pm} ({\sf \Sigma}'')_{s,\overline s}^{\lambda,\eta}(\epsilon)
                             \,   \sin \left ( 2\pi \xi_{\overline s}(0) \right )            \nonumber,
\end{align}
where ``+ \dots'' contains the non-oscillatory interaction self-energy contribution, the term (arising from disorder scattering) which oscillates at the frequency as the $s$ band,
as well as terms with higher powers of the Dingle factor.

\subsubsection{Difference frequency oscillations}
\label{sec:diff-freq-oscill}

Plugging Eq.\eqref{eq:29} into Eq.\eqref{eq:19} one obtains the difference frequency contribution 
\begin{align}
  \label{eq:35}
  \tilde \rho_{\rm df}(\epsilon)
  &\approx  N_\phi\sum_{s}\frac {1} { \omega_{{\rm c},s}^2 }  \, R_{\rm D}^{ s}\,
         \Big (  \exp\left  (i 2\pi \left (  \xi_{ s}(\epsilon) -\xi_{\overline s} (0)  \right )\right ) \sum_\lambda \sum_{\eta=\pm} ( {\sf \Sigma}'')_{s,\overline s}^{\lambda,\eta}(\epsilon) \;+{\rm h.c.} \Big ) .
\end{align}

We now compute the energy integral Eq.\eqref{eq:7}.
In Eq.\eqref{eq:77b} we take $\xi_s(\epsilon) \rightarrow \xi_s(0)$,
so that $\sum_{\eta=\pm} ({\sf \Sigma}'')_{s,\overline s}^{\lambda,\eta}(\epsilon)
= (\hat {\sf \Sigma}'')_{s,\overline s}^{\lambda}\sum_{\eta=\pm}\eta(\epsilon+\eta \bar \Omega_\lambda + i \Gamma_{\overline s}^{\rm dis} )^{-1}$,
and the difference frequency oscillatory term in $n_{\rm p}$ reads
\begin{align}
  \label{eq:80}
  \tilde n_{\rm p}^{\rm df}
  &\approx  {\rm Re}\sum_{s,\lambda} \sum_{\eta=\pm} \frac {1} { \omega_{{\rm c},s}^2 }  (\hat {\sf \Sigma}'')_{s,\overline s}^{\lambda} \, R_{\rm D}^{s}
  \,  e^{ i 2\pi \left (\xi_{s}(0) - \xi_{\overline s} (0) \right )}
    \int \d \epsilon \, n_{\rm F}(\epsilon) \,e^{i 2\pi m_s  \epsilon } \, \frac \eta {\epsilon+\eta \bar \Omega_\lambda + i \Gamma_{\overline s}^{\rm dis} } ,\\
    (\hat {\sf \Sigma}'')_{s,\overline s}^{\lambda}
    &= - \frac 1 \pi \frac{2\pi^2 T /\omega_{{\rm c},\overline s}}{\sinh(2\pi^2T/\omega_{{\rm c},\overline s} )}
  \, (2\pi/ \bar \Omega_{\lambda}) \,\times \mc I_{\xi_s(0), \xi_s(0)}^{\lambda;s;0}(0).
\end{align}

The energy integral is
\begin{align}
  \label{eq:49}
\int \d \epsilon \, n_{\rm F}(\epsilon) \,e^{i 2\pi m_s  \epsilon } \,  \sum_{\eta=\pm}  \frac \eta {\epsilon+\eta \bar \Omega_\lambda + i \Gamma_{\overline s}^{\rm dis} }
  &= 2\pi T \sum_{{\rm n}\geq 0} e^{-(2{\rm n}+1)2\pi^2 m_s T} \sum_{\eta=\pm}  \frac \eta {(2{\rm n}+1)\pi T + \Gamma_{\overline s}^{\rm dis}+ i \eta \bar \Omega_\lambda  }.
\end{align}

In the limit of a small Dingle factor, $\Gamma_{\overline s}^{\rm dis} /\omega_{{\rm c},s} \gtrsim 1$, one can use the first Matsubara frequency approximation
(i.e.\ neglect all ${\rm n}\geq 1$ dependence outside of the exponential thermal factor). This yields Eq.\eqref{eq:main76} in the main text, where
\begin{align}
  \label{eq:31}
  \ms A_{n_{\rm p}}^{s,\lambda}(T)  &= -\frac 1 {\pi}  \, \frac 1 {\bar \Omega_{\lambda}} \frac {1} { \omega_{{\rm c},s} } \, e^{-2\pi \Gamma_{\overline s}^{\rm dis}/\omega_{{\rm c},s}}  \,
                                    R_{\rm LK}^{[m_s]}(T)\,   R_{\rm LK}^{[m_{\overline s}]}(T)\,
                                       \frac{2 \bar \Omega_\lambda }{\bar \Omega_\lambda ^2 + (\Gamma_{\overline s}^{\rm dis}+\pi T)^2}
 \,\times \mc I_{\xi_s(0), \xi_s(0)}^{\lambda;s;0}(0) .
\end{align}

\subsubsection{Sum frequency oscillations}
\label{sec:sum-freq-oscill}

Because the main dHvA difference frequency oscillation decays as rapidly as the sum frequency when increasing temperature, we also compute the leading two sum frequency terms:
\begin{align}
  \label{eq:11}
 \tilde \rho_{\rm sf}(\epsilon)
  &\approx  - N_\phi\sum_{s}\frac {1} { \omega_{{\rm c},s}^2 }  \, R_{\rm D}^{ s}\,
    \sum_\lambda \sum_{\eta=\pm} \Big (  e^{i 2\pi \left (  \xi_{ s}(\epsilon) +\xi_{\overline s} (0)  \right )}  ( {\sf \Sigma}'')_{s,\overline s}^{\lambda,\eta}(\epsilon)
    -  e^{i 2\pi \left (  \xi_{ s}(\epsilon) +\xi_{\lambda,\overline s}^\eta (\epsilon)  \right )} i ( {\sf \Sigma}')_{s,\overline s}^{\lambda,\eta}(\epsilon)   \Big ) \;+{\rm h.c.} 
\end{align}

With $ \sum_{\eta=\pm} ({\sf \Sigma}'')_{s,\overline s}^{\lambda,\eta}(\epsilon)
= (\hat {\sf \Sigma}'')_{s,\overline s}^{\lambda}\sum_{\eta=\pm}\eta(\epsilon+\eta \bar \Omega_\lambda + i \Gamma_{\overline s}^{\rm dis} )^{-1}$
and $({\sf \Sigma}')_{s,\overline s}^{\lambda,\eta}(\epsilon)
= \left [ n_{\rm B}(\bar \Omega_{\lambda}) + n_{\rm  F}( \eta \epsilon + \bar \Omega_{\lambda} ) \right ] \, (\hat{\sf \Sigma}')_{s,\overline s}^{\lambda}$,
we evaluate the energy integrals using Eq.\eqref{eq:49} and the integrals
\begin{align}
  \label{eq:49a}
\int_{-\infty}^{+\infty}\d \epsilon  \,n_{\rm F}(\epsilon) \,  e^{i X\epsilon}
  &= \frac{-2i \pi T}{2\sinh(\pi X T)} ,\\
    \label{eq:49b}
   \int_{-\infty}^{+\infty}\d \epsilon  \,n_{\rm F}(\epsilon) \, n_{\rm F}(\eta \epsilon + \Omega ) \, e^{i X \epsilon}
  &= \frac{+2i \pi T}{2\sinh(\pi X T)} \left (  n_{\rm B}(\Omega) + \eta\, n_{\rm B}(-\eta \Omega)\,e^{-i \eta \Omega X} \right ),
\end{align}
which together with $ \eta\, n_{\rm B}(-\eta \Omega) = - \left ( n_{\rm B}(\Omega)+\tfrac{1+\eta}2 \right )$ yields
\begin{align}
  \label{eq:15}
   \tilde n_{\rm p}^{\rm sf} &\approx \sum_{s,\lambda}\, (\ms A')_{n_{\rm p}}^{s,\lambda}(T) \, \sum_{\eta = \pm}
                               \exp \left ( \tfrac{1+\eta}2  \bar \Omega_{\lambda}/T \right )\, \cos \left ( 2\pi (\xi_s(0) + \xi_{\overline s}(0) - \eta \bar \Omega_{\lambda}/\omega_{{\rm c},s} ) \right )  \nonumber \\
  & +  \sum_{s,\lambda}\, (\ms A'')_{n_{\rm p}}^{s,\lambda}(T) \, \sin \left ( 2\pi (\xi_s(0) + \xi_{\overline s}(0) )\right )  ,\\
  (\ms A')_{n_{\rm p}}^{s,\lambda}(T)&\equiv  \frac 1 {2\pi^2} \frac {1} { \omega_{{\rm c},s}^2 }  \, R_{\rm D}^{ s}\,  (\hat {\sf \Sigma}')_{s,\overline s}^{\lambda}
                                       \, n_{\rm B}(\bar \Omega_{\lambda}) \;  \frac{2\pi^2 T}{\sinh\left (2\pi^2 (m_s+m_{\overline s})T \right )}, \\
  (\ms A'')_{n_{\rm p}}^{s,\lambda}(T)&\equiv - \frac 1 {2\pi^2} \frac {1} { \omega_{{\rm c},s} }  \, ^{-2\pi \Gamma_{\overline s}^{\rm dis}/\omega_{{\rm c},s}}  \,  (\hat {\sf \Sigma}'')_{s,\overline s}^{\lambda}
                                       \; \frac{2 \bar \Omega_\lambda }{\bar \Omega_\lambda ^2 + (\Gamma_{\overline s}^{\rm dis}+\pi T)^2}\,   R_{\rm LK}^{[m_s]}(T).
\end{align}

When $T_b\neq T_f$, a new term appears at the same sum frequencies,
from the absence of cancellation in $n_{\rm B}(\bar \Omega_\lambda) \text{Eq.\eqref{eq:49a}} + \text{Eq.\eqref{eq:49b}}$:
\begin{align}
  \label{eq:82}
  \tilde n_{\rm p}^{\rm sf,2T} &\approx  \sum_{s,\lambda} \ms A_{n_{\rm p},\rm 2T}^{s,\lambda}(T_b,T)  \sum_{\eta=\pm}\,  
                                 \cos \left (  2\pi ( \xi_{\lambda,\overline s}^\eta(0) + \xi_{s}(0) ) \right ) ,\\
\ms A_{n_{\rm p},\rm 2T}^{s,\lambda}(T_b,T) &\equiv (\ms A')_{n_{\rm p}}^{s,\lambda}(T) \, \left ( n^{T_b}_{\rm B}(\bar \Omega_{\lambda}) - n^{T}_{\rm B}(\bar \Omega_{\lambda}) \right ).
\end{align}

\end{widetext}
     
\end{document}